\documentclass{aa}  

\usepackage{graphicx}
\usepackage{txfonts}
\usepackage{lscape}
\usepackage{colortbl} 
\usepackage[dvipsnames]{xcolor}
\usepackage{epstopdf}
\usepackage{ulem}
\usepackage{diagbox}
\begin{document}

\title{Lithium depletion and angular momentum transport in solar-type stars}

   \authorrunning{T. Dumont et al.}
   
   \titlerunning{Lithium depletion and angular momentum transport in solar-type stars}
   

   \author{T. Dumont
          \inst{1,2} \fnmsep\thanks{e-mail: thibaut.dumont@unige.ch},
          A. Palacios \inst{2},
          C. Charbonnel \inst{1,3},
          O. Richard \inst{2},
          L. Amard \inst{4},
          K. Augustson \inst{5}
                    \and
          S. Mathis \inst{5}
          }

   \institute{Department of Astronomy, University of Geneva, Chemin des Maillettes 51, 1290 Versoix, Switzerland
         \and LUPM, Universit\'e de Montpellier, CNRS, Place Eug\`ene Bataillon, 34095 Montpellier, France
         \and IRAP, UMR 5277 \& Universit\'e de Toulouse, 14 avenue Edouard Belin, 31400 Toulouse, France
         \and University of Exeter, Department of Physics and Astronomy, Stocker Road, Devon, Exeter, EX4 4QL, United Kingdom
         \and AIM, CEA, CNRS, Universit\'e Paris-Saclay, Universit\'e Paris-Diderot, Sorbonne Paris Cit\'e, F-91191 Gif-sur-Yvette Cedex, France
             }

   \date{(Received; Revised; Accepted)}

 
  \abstract
   {Transport processes occurring in the radiative interior of solar-type stars are evidenced by the surface variation of light elements, in particular $^{7}{\rm Li}$, and the evolution of their rotation rates.
   For the Sun, inversions of helioseismic data indicate that the radial profile of angular velocity in its radiative zone is nearly uniform, which implies the existence of angular momentum transport mechanisms that are efficient over evolutionary timescales. While there are many independent transport models for angular momentum and chemical species, there is a lack of self-consistent theories that permit stellar evolution models to simultaneously match the present-day observations of solar lithium abundances and radial rotation profiles.}
   {We explore how additional transport processes can improve the agreement between evolutionary models of rotating stars and observations for $^{7}{\rm Li}$ depletion, the rotation evolution of solar-type stars, and the solar rotation profile.
   }
   {Models of solar-type stars are computed including atomic diffusion and rotation-induced mixing with the code STAREVOL. We explore different additional transport processes for  chemicals and for angular momentum such as penetrative convection, tachocline mixing, and additional turbulence. We constrain the resulting models by  simultaneously using the evolution of the surface rotation rate and $^{7}{\rm Li}$ abundance in the solar-type stars of open clusters with different ages, and the solar surface and internal rotation profile as inverted from helioseismology when our models reach the age of the Sun.
   }
   {We show the relevance of penetrative convection for the depletion of $^{7}{\rm Li}$ in pre-main sequence and early main sequence stars. The rotational dependence of the depth of penetrative convection yields an anti-correlation between the initial rotation rate and $^{7}{\rm Li}$ depletion in our models of solar-type stars that is in agreement with the observed trend. Simultaneously, the addition of an ad hoc vertical viscosity $\rm{\nu_{add}}$ leads to efficient transport of angular momentum between the core and the envelope during the main sequence evolution and to solar-type models that match the observed profile of the Sun. We also self-consistently compute for the first time the thickness of the tachocline and find that it is compatible with helioseismic estimations at the age of the Sun, but we highlight that the associated turbulence does not allow  the observed $^{7}{\rm Li}$ depletion to be reproduced. The main sequence depletion of $^{7}{\rm Li}$ in solar-type stars is only reproduced when adding a parametric turbulent mixing below the convective envelope.
   }
   {The need for additional transport processes in stellar evolution models for both chemicals and angular momentum in addition to atomic diffusion, meridional circulation, and turbulent shear is confirmed. We identify the rotational dependence of the penetrative convection as a key process. Two  additional and distinct parametric turbulent mixing processes (one for angular momentum and one for chemicals) are required to  simultaneously explain the observed surface $^{7}{\rm Li}$ depletion and the solar internal rotation profile. We highlight the need of additional constraints for the internal rotation of young solar-type stars and also for the beryllium abundances of open clusters in order to test our predictions.  
   }


   \maketitle
   
%

\section{Introduction}
\label{section:introduction}

   Solar-type stars (e.g. stars with an initial mass of 1$\pm 0.1$ M$_\odot$ and a value of [Fe/H] = 0$\pm$0.2~dex, but possibly not the same age as the Sun) have been extensively used to constrain and study transport processes of chemicals and angular momentum in stellar interiors \citep[e.g.][]{1989ApJ...338..424P,1992A&A...265..115Z,1996A&A...312.1000R,2002ApJ...574L.175T,2003A&A...405.1025T,2005A&A...440..981T,2005A&A...440L...9E,2019A&A...626L...1E,2008IAUS..252..163C,2009A&A...494..663C,2016A&A...587A.105A,2017ApJ...845L...6B}. 
   Currently the main challenge is to  simultaneously explain the time evolution of their photospheric $\rm ^7Li$ (hereafter Li) abundances, and of the rotation of their surface and interior.  
  Several hints point to rotation-induced transport and mixing processes as the cause for Li depletion in the Sun and in main sequence (MS) solar-type stars \citep[e.g.][]{1987A&A...175...99L,1990ApJS...74..501P,1992A&A...255..191C,1996A&A...312.1000R,2009A&A...501..687D}, although other mechanisms have been invoked, such as penetrative convection (\citealt[][]{1963ApJ...138..297B}, \citealt[][]{1999A&A...347..272S}, \citealt[][]{2017ApJ...845L...6B}, \citealt[][]{2018MNRAS.481.4389J}); tachocline mixing (\citealt{1999ApJ...525.1032B}); mass loss (\citealt{2010ApJ...713.1108G}); planet accretion (\citealt{2002A&A...386.1039M}); and mixing by internal gravity waves (\citealt{1994A&A...281..421M}). 
  On the other hand, the currently available prescriptions for anisotropic turbulence and meridional circulation that are induced by rotation and that transport both chemicals and angular momentum fail to reproduce the internal rotation rates evidenced by helio- and asteroseismology \citep[e.g.][]{2012A&A...548A..10M,2012ApJ...756...19D,2014A&A...564A..27D,2015A&A...580A..96D,2020arXiv200702585D,2013A&A...555A..54C,2013A&A...549A..74M,2015MNRAS.452.2654B,2017A&A...599A..18E,2019A&A...626L...1E,2018A&A...616A..24G,2019LRSP...16....4G,2018A&A...620A..22M,2019A&A...631A..77A,2019ARA&A..57...35A}.
  More efficient mechanisms are required for the transport of angular momentum, which could be driven by internal gravity waves or magnetic processes and instabilities \citep[e.g.][]{1993A&A...279..431S,2002A&A...381..923S,2005Sci...309.2189C,2005A&A...440..653M,2005A&A...440L...9E,2010A&A...519A.116E,2019A&A...621A..66E,2019A&A...631L...6E,2010ApJ...716.1269D,2013A&A...554A..40C,2015A&A...579A..31B,2017A&A...605A..31P,2019MNRAS.485.3661F}. These mechanisms also impact the transport of chemicals induced by rotation and influence the way Li is depleted with time in solar-type stars \citep[e.g.][]{2005Sci...309.2189C,2005A&A...440..981T}.
  
   In this work we explore the possibilities to reproduce simultaneously the chemical and rotational constraints for solar-type stars along their evolution up to the end of the MS.  
In \S~\ref{sec:obs} we present the observational data that we aim to account for with our stellar evolution models.  
In \S~\ref{STAREVOL} we describe the input physics of the models and recall the state-of-the-art expressions from the literature for the different transport processes implemented in the stellar evolution code STAREVOL, and tested in this work. 
In \S~\ref{MODELS} we compare the predictions of the so-called Type I models for rotating stars (which only include  meridional circulation, shear induced turbulence, and atomic diffusion) to the observational constraints for solar-type stars over a broad age range. In \S~\ref{sec:improved} we probe for effects of penetrative convection and tachocline turbulence using for the first time some rotation-dependent prescriptions.  
This allows us to quantify the efficiency of still missing processes 
that we simulate in the form of a vertical diffusivity for the transport of angular momentum, and of an ad hoc turbulence for the transport of chemicals.
We summarise our results and conclude in \S~\ref{CONCLUSION}.
  
\section{Observational constraints}
\label{sec:obs}
To best constrain the physics at play in the interior of solar-type stars along their evolution up to the MS turnoff, we use both chemical and rotational data.

\subsection{Lithium abundances}
\label{sub:Lisuncluster}

Lithium has long been claimed to be a useful and constraining element that can be used to understand the  transport of chemicals and of angular momentum in stellar interiors \citep[e.g.][]{1969ARA&A...7...99W,1976PASP...88..353B,1978ApJ...223..567V,1982A&A...115..357S,1985A&A...149..309B,1988ApJ...335..971V,1987A&A...175...99L,1990ASSL..159..437B,1991ApJ...370L..95B,1992A&A...255..191C,1994A&A...283..155C,2000IAUS..198...61D,1996A&A...305..513M,2000A&A...354..943M,2002ApJ...566..419P,2005Sci...309.2189C,2010IAUS..268..365T}. 
   Because of its relatively low burning temperature ($\sim 2.5$~MK, close to the temperature at the base of the convective zone in solar-type stars),
   Li is indeed easily destroyed by proton capture in stellar interiors. According to classical stellar evolution theory, this destruction is expected to manifest itself at the surface of solar-type stars during the pre-main sequence (PMS) in the form of a decrease in  the surface Li abundance. Classical models that include no transport processes beyond convection predict no further surface Li variation until the first dredge-up episode when the stars evolve towards the red giant branch.  
   
   Spectroscopic observations, however, show that the abundance of lithium at the surface of field and open cluster solar-type stars decreases along the main sequence \citep[e.g.][]{1997AJ....113.1871K,2005A&A...442..615S,2006AJ....131.1816C,2010A&A...515A..93T,2011A&A...535A..75S,2012A&A...537A..91X,2014A&A...562A..92D,2017MNRAS.465.2076W,2017AJ....153..128C,2017A&A...602A..63B,2018A&A...618A..16H,2020MNRAS.492..245C}. In the case of the Sun, Li has decreased from its original  (i.e. meteoritic) value of $A(^7Li) = 3.31$\footnote{$A(X) = log_{10}(N_{X}/N_H)+12$ (where $N_X$ is the number density of element X)} down to  $A(^7Li)=1.05$  \citep{1951ApJ...113..536G,1957ApJ...125..233S,2009ARA&A..47..481A}. 
 Solar twins (solar-type stars with ages close to that of the Sun, i.e. 4.6$\pm 0.5$~Gyr) all present significant Li depletion, with non-negligible dispersion, and with the Sun being among the most Li-depleted \citep[e.g.][]{2007A&A...468..663T,2007ApJ...669L..89M,2019MNRAS.485.4052C}.
 
In this work we use a consistent set of lithium abundances provided by \citet{2005A&A...442..615S} for a group of open clusters with [Fe/H] between -0.21 and +0.14~dex: NGC 2264, IC2391, IC2602, IC4665, $\alpha$ Per, Pleiades, Blanco I, NGC2516, M34, NGC6475, M35, Praesepe, NGC6633, and NGC752. We identify the solar-type stars as those that have an effective temperature as derived by \citet{2005A&A...442..615S}, corresponding to the effective temperature $\pm$100 K of our model including atomic diffusion and rotation at the age of the corresponding cluster. 
We assume the cluster ages given by \citet{2019A&A...623A.108B}. Given the relatively large uncertainty on actual age determination, this slight inconsistency with the ages that would be derived with our models should not affect our conclusions.
In Fig.~\ref{fig:evolLi1} (also appearing in Figs.~\ref{fig:evolLi2}, \ref{fig:evolLi3}, \ref{fig:evolLi4}, and \ref{fig:evolLi5}) we show the corresponding Li range for the solar-type stars with the observational boxes, and indicate  the age uncertainty for each cluster.
We also consider and show the Li data for M67 and field solar twins by \citet{2019MNRAS.485.4052C,2020MNRAS.492..245C}.

\subsection{Surface and internal rotation} 
\label{subsub:rotationprofile}
\subsubsection{Surface rotation}
Many observations exist of the rotation rates of solar-type stars of different ages \citep[e.g.][]{1986PASP...98.1233S,2014ApJS..211...24M,2014A&A...572A..34G,2015A&A...577A..98G,2016A&A...592A.156D,2020MNRAS.tmpL..51L}. They clearly establish  that the surface rotation of these stars evolves with time under the effect of multiple processes. Magnetic interactions between the star and its accretion disc early on the PMS and later with its wind are successfully invoked to explain the evolution and the dispersion of the rotation periods provided by photometric surveys \citep[e.g.][and references therein]{2015ApJ...799L..23M,2016A&A...587A.105A,2019A&A...632A...6G}. 

To constrain the surface rotation of the models we use the observational data set gathered by \citet{2015A&A...577A..98G} for a large number of solar-type stars in open clusters of various ages. We also use the data from \citet{2016ApJ...823...16B} for a subsample of four stars of M67, selected because their magnitudes are close to the solar value, with a magnitude $B-V \in [0.6;0.7]$.\\ 

\subsubsection{Internal rotation}
The internal rotation of the Sun is constrained by helioseismology. It is assumed to be similar for other solar-type stars, and asteroseismology constrains it for stars in more advanced evolutionary stages beyond the MS \citep[i.e. subgiant and red giant stars; e.g.][]{2012A&A...548A..10M,2012ApJ...756...19D,2014A&A...564A..27D,2015A&A...580A..96D,2015MNRAS.452.2654B,2018A&A...616A..24G,2019LRSP...16....4G}. 
The analysis of p-modes gives access to the rotation profile of the Sun between about $R=0.2 R_{\odot}$ and the surface  \citep{1988SvAL...14..145K,1995Natur.376..669E,2003ARA&A..41..599T,2008A&A...484..517M,2008ApJ...679.1636E}. The inverted rotation profile is compatible with solid-body rotation in the radiative zone. Moreover, the analysis of mixed modes in solar-mass subgiant and red giant stars (hereafter SGB and RGB) presenting solar-like oscillations also points to a low degree of radial differential rotation in the core indicating that the strong coupling found at the solar age is essentially maintained during further evolution \citep{2017A&A...599A..18E,2019A&A...626L...1E,2018A&A...620A..22M,2019ARA&A..57...35A}.
No observational clue exists yet regarding the structure of the internal rotation during the PMS and the early MS evolution, although there are hints that the quasi  solid-body rotation of the solar interior may not be an exception from the analysis of asteroseismic data for solar-type stars \citep{2014A&A...568L..12N} and for F to late G main sequence stars \citep{2015MNRAS.452.2654B}. 

We thus require that our best models reach an internal rotation profile similar to the solar profile at the age of the Sun as this is basically the only proper constraint for the phases investigated in this paper.  For this profile we use the results of the inversion of MDI-GOLF-GONG\footnote{MDI: Michelson Doppler Imager \citep{1995SoPh..162..129S}, GOLF: Global Oscillations at Low Frequencies \citep{1995SoPh..162...61G}. These instruments are on board the SOlar and Heliospheric Observatory (SOHO) spacecraft of ESA/NASA \citep{1995SoPh..162....1D}. GONG:  Global Oscillation Network Group \citep[e.g.][and references therein]{2020MNRAS.493L..49H}.} data by \citet{2008ApJ...679.1636E}.

\section{Stellar evolution models}
\label{STAREVOL}

We use an updated version of the stellar evolution code STAREVOL \citep[for general information and previous versions, see ][]{2000A&A...358..593S,2006A&A...453..261P,2009A&A...495..271D,2012A&A...543A.108L,2019A&A...631A..77A}. All our models are evolved without accretion starting prior to the deuterium birthline on the PMS from initial structures corresponding to homogeneous polytropes. This sets the time zero of our computations.

\subsection{Input physics}
\label{sub:inputphysics}
We adopt the solar reference abundances from \citet{2009ARA&A..47..481A} including the enhancement of neon recommended by \citet{2018ApJ...855...15Y} as reported in Table \ref{abondini}. The opacities are interpolated within the OPAL opacity tables \citep{1996ApJ...464..943I} when T > 8000 K and the low-temperature opacity tables from the Wichita opacity database when T < 8000 K (Ferguson, J. W., private communication) that are fully consistent with the adopted solar reference abundances. 
\begin{table}[t]
\caption{Initial chemical mixture adopted for the different solar calibrated models 
}
   \centering
    \begin{tabular}{ccc}
        \hline \hline
        Model & C & R \\
        Element & & \\
        \hline
         $^1H$ & $7.18\,\times\,10^{-1}$ & $7.14\,\times\,10^{-1}$ \\
         $^4He$ & $2.69\,\times\,10^{-1}$ & $2.72\,\times\,10^{-1}$ \\
         $^7Li$\tablefootmark{*} & $1.03\,\times\,10^{-8}$ & $1.03\,\times\,10^{-8}$ \\
         $^9Be$\tablefootmark{$\dagger$} & $1.66\,\times\,10^{-10}$ & $1.65\,\times\,10^{-10}$ \\
         $^{11}B$ & $3.13\,\times\,10^{-9}$ & $3.31\,\times\,10^{-9}$ \\
         $^{12}C$ & $2.26\,\times\,10^{-3}$ & $2.40\,\times\,10^{-3}$ \\
         $^{14}N$ & $6.68\,\times\,10^{-4}$ & $7.07\,\times\,10^{-4}$ \\
         $^{16}O$ & $5.54\,\times\,10^{-3}$ & $5.86\,\times\,10^{-3}$ \\
         $^{19}F$ & $4.89\,\times\,10^{-7}$ & $5.17\,\times\,10^{-7}$ \\
         $^{20}Ne$ & $1.58\,\times\,10^{-3}$ & $1.68\,\times\,10^{-3}$ \\
         $^{23}Na$ & $2.83\,\times\,10^{-5}$ & $3.00\,\times\,10^{-5}$ \\
         $^{24}Mg$ & $5.35\,\times\,10^{-4}$ & $5.66\,\times\,10^{-4}$ \\
         $^{27}Al$ & $5.39\,\times\,10^{-5}$ & $5.71\,\times\,10^{-5}$ \\
         $^{28}Si$ & $5.92\,\times\,10^{-4}$ & $6.27\,\times\,10^{-4}$ \\
         $^{31}P$ & $5.65\,\times\,10^{-6}$ & $5.97\,\times\,10^{-6}$ \\
         $^{32}S$ & $2.84\,\times\,10^{-4}$ & $3.01\,\times\,10^{-4}$ \\
         $^{35}Cl$ & $6.73\,\times\,10^{-8}$ & $7.12\,\times\,10^{-8}$\\
         $\rm{Others}$\tablefootmark{**} & $1.49\,\times\,10^{-3}$ & $1.58\,\times\,10^{-3}$ \\
        \hline
        \label{table_1}
    \end{tabular}
    \tablefoot{Initial abundances are given in mass fraction for the classical model (C) and the rotation model (R). Each model has been calibrated on the Sun (see Table \ref{table_calib}). The mass fraction of metals $Z$ is $Z_C = 0.0134$ and $Z_R = 0.0142$ for the classical and rotating calibrated solar models, respectively.\\
    \tablefoottext{*} $A_{ini}(^7Li) = 3.31$\\
    \tablefoottext{$\dagger$} $A_{ini}(^9Be) = 1.41$ \\
    \tablefoottext{**} \,refers to elements heavier than chlorine.}
    \label{abondini}
\end{table}{}

The equation of state is analytical and follows \citet{1973A&A....23..325E} and \citet{1995MNRAS.274..964P}, as described in \citet{2000A&A...358..593S}.
We use the nuclear reactions rates from the NACRE2 database
generated using the NetGen web interface \citep{2013A&A...549A.106X,2013NuPhA.918...61X}. 
\begin{table*}[t!]
  \centering
 \caption{Solar calibration results for the STAREVOL classical model (C: no transport other than convection) and rotating model (R:  atomic diffusion and type I rotation-induced transport).} 
 \begin{tabular}{l c c c}
    \hline
    \hline
    \small
      & Sun & Model C &  Model R \\
    \hline
     Y$_{\rm{surf}}$ & 0.2485 $^1$ & 0.2685 & 0.2559  \\
     Z$_{\rm{surf}}$ & 0.0134 $^2$ & 0.0134 & 0.0134  \\
     $\frac{Z_{\rm{surf}}}{X_{\rm{surf}}}$ & 0.0181 $^2$ & 0.0186 & 0.0183 \\
     T$_{\rm{eff}} (\rm{K})$ & 5777 & 5775 & 5779 \\
     L$_\odot ~(10^{33} \rm{erg.s^{-1}})$  & 3.846 $^3$ & 3.846 & 3.845 \\
     R$_\odot ~(10^{10} \rm{cm})$ & 6.9599 $^4$ & 6.9599 & 6.9555 \\
     \hline
     Relative luminosity accuracy: dL  & $\dots$ & $10^{-6}$ & 3.0 $\times 10^{-4}$ \\
     Relative radius accuracy: dR & $\dots$& $10^{-6}$ & 6.3 $\times 10^{-4}$\\ 
     \hline
     $Y_{\rm{ini}}$ & $\dots$& 0.2685 & 0.2718 \\
     $Z_{\rm{ini}}$ & $\dots$& 0.0134 & 0.0142  \\
     $\alpha_{\rm{MLT}}$ & $\dots$& 2.110 & 2.223  \\
     \hline
     \label{table_calib}
  \end{tabular}
 
  \tablefoot{
  \newline
  $Y_{\rm{surf}}$ and $Z_{\rm{surf}}$ are respectively the surface helium and heavy element mass fraction, $\frac{Z_{\rm{surf}}}{X_{\rm{surf}}}$ is the ratio of heavy element to the hydrogen mass fractions, T$_{\rm{eff}}$ is the effective temperature (K), L$_\odot$ is the luminosity (in solar units), R$_\odot$ is the radius (in solar units), $Y_{\rm{ini}}$ and $Z_{\rm{ini}}$ are respectively the initial helium and heavy element mass fractions, and $\alpha_{\rm{MLT}}$ is the mixing length parameter.\\
  $^1$ Helioseismic estimation in the convective zone from \citet{1995MNRAS.276.1402B} \\
  $^2$ \citet{2009ARA&A..47..481A}\\ 
  $^3$ \citet{1995RvMP...67..781B} \\ 
  $^4$ \citet{1976asqu.book.....A}\\
  }
  \label{tab:solcal}
\end{table*}

In the current version of STAREVOL the full set of stellar structure equations is solved for the whole star;  there is no decoupling between the interior and the envelope (where the diffusion approximation becomes valid). The surface boundary conditions are treated using the  Hopf function $q(\tau)$, which provides a correction to the grey approximation (see \citealt{1930MNRAS..90..287H,1994A&A...286...91M}) 
\begin{equation}
\frac{4}{3}\left(\frac{T(\tau)}{T_{\rm{eff}}}\right)^4 = q(\tau) + \tau
\label{Eq:qtau}
\end{equation}
at a given optical depth $\tau$, $T_{\rm eff}$ being the temperature of the equivalent black body and $T(\tau)$ the temperature profile. We use the analytical expression from \citet{1966ApJ...145..174K} for $q(\tau)$, as is also done in \citet{2013A&A...558A..46P} and \citet{2019ApJ...881..103Z}, for instance. The numerical surface is set at $\tau_0 = 0.005$, as in \citet{2019A&A...631A..77A}, and the connection to the atmosphere is made at $\tau_{ph} = 2$. 
\newline
The models without rotation take into account mass loss starting at the ZAMS\footnote{Zero Age Main Sequence} following the empirical relation by \citet{1975MSRSL...8..369R}, with $\eta_{R} = 0.5$, as advocated by \citet{2015MNRAS.448..502M} and \citet{2017Ap&SS.362...15G} for solar-type stars from observational constraints on the red giant branch. When the effects of rotation are taken into account, we use the mass loss prescription by \citet{2011ApJ...741...54C}, as in \citet{2019A&A...631A..77A}. 

Heat transport by convection follows the Mixing Length Theory \citep[MLT;][]{1958ZA.....46..108B,1968pss..book.....C}. 
The convective boundaries are determined with the Schwarzschild criterion. When included, the effect of penetrative convection is treated as overshoot \citep[without changing the temperature gradient in the concerned region; see][]{1991A&A...252..179Z} below the convective envelope. 

\subsection{Model calibration}
The abundances of helium and metals vary according to the input physics of the models; in other words,  they depend on the mixing processes considered. Consequently, for each case the initial chemical composition needs to be evaluated so that the ratio  $Z_{\rm surf} / X_{\rm surf}$  from \citet[][]{2009ARA&A..47..481A} is reproduced at the age of the Sun. In this calibration procedure the mixing length parameter $\alpha_{\rm MLT}$ and the initial chemical composition are calibrated so as to reproduce the solar radius and solar luminosity at the age of the Sun (4.57 Gyr) with a relative accuracy of the order of $10^{-4}$ to $10^{-6}$. We make two different calibrations depending on the physics of the models (see Table~\ref{tab:solcal}). The classical model (C) is without any transport processes in the radiative region and the rotating model (R) includes atomic diffusion and rotational mixing with the assumptions made for model R1 (Table~\ref{tab:allparam}) and the median rotation rate (see \S~\ref{sub:stellarrotation}), but it does not include overshoot. The calibration corresponding to model R1 is then used for {all} the models produced including rotation, in particular for models $^a_b$R$_c^d$ which include additional transport processes for angular momentum and chemicals (see \S~\ref{MODELS} and
the Appendix for further details). Details on the models resulting from these calibrations are given in Table~\ref{tab:solcal} and the initial chemical mixtures for each calibration are reported in Table \ref{abondini}.\\ 

\subsection{Evolution of chemical abundances:  General equation and atomic diffusion}
\label{sub:atomicdiffusion}

Chemical abundances within the star evolve under the effect of nuclear reactions and transport processes. This is described by the general diffusion equation \citep[e.g.][]{2009pfer.book.....M}, which involves the different physical processes operating in the star
\begin{equation}
    \rho \frac{\partial X_i}{\partial t} = \frac{1}{r^2}\frac{\partial}{\partial r} \left(r^2 \rho D \frac{\partial X_i}{\partial r}\right)-\frac{1}{r^2}\frac{\partial}{\partial r} \left(r^2 \rho X_i \rm{v_i}\right)+ m_i \left[ \sum_j r_{ji} - \sum_k r_{ik} \right],
    \label{eqdiffu}
\end{equation}
where $\rho$ is the density; $X_i$ refers to the mass fraction of element $i$; $r$ is the radius; $D = \sum_j D_j$ is the total coefficient for turbulent diffusion, written as the sum of the $j$ different diffusion coefficients describing turbulent processes such as shear, penetrative convection, or any other unidentified process (see Sections~\ref{subsub:rotationprescriptions}, \ref{sub:penetrativeconvection}, and \ref{sub:addchemicaltransp}); $\rm{v_i}$ is the diffusion velocity of element $i$; $m_i$ is the mass of nuclei $i$; and $r_{ij}$ the reaction rate producing nuclei $j$ from nuclei $i$.

Atomic diffusion is implemented in STAREVOL with the formalism of \citet{1994ApJ...421..828T} to solve the Burgers equations and compute the individual atomic diffusion velocities of each element taken into account in STAREVOL (see Table~\ref{table_1}). The computation of the collision integrals is done according to \citet{1986ApJS...61..177P}. We take into account the partial ionisation of chemical elements for temperatures lower than $5\times10^6$K \citep{2002A&A...395...85S}. 
Radiative accelerations are not taken into account in our models. According to \citet{1998ApJ...504..539T} their impact on abundances for light elements in the solar case is only about $2\%$. Radiative accelerations mainly impact the heavy elements such as iron, and become important for stars more massive than solar-type stars \citep{1998ApJ...492..833R,2002ApJ...568..979R,2018A&A...618A..10D}.

\subsection{Angular momentum evolution and rotation-induced mixing}
\label{sub:stellarrotation}
Stellar rotation, and in particular differential rotation, is a potent trigger of transport for both angular momentum and chemicals in stellar radiation zones. It generates the large-scale currents of the meridional circulation and several large-scale hydrodynamical instabilities that induce turbulence such as the vertical and horizontal shear instabilities, which are the ones  included in our models \citep[][]{1992A&A...265..115Z,1998A&A...334.1000M,2004A&A...425..229M,2004A&A...425..243M}. This ensemble is referred to as Type I rotational mixing as it does not include  the transport by magnetohydrodynamic (MHD) instabilities and magnetic fields or transport by internal gravity waves (see e.g. \citealt{2013LNP...865...23M,2019ARA&A..57...35A} for  a description of these processes).

\subsubsection{Prescriptions for shear induced turbulent transport}
\label{subsub:rotationprescriptions}
Stellar rotation is implemented in STAREVOL as described by \citet{2016A&A...587A.105A,2019A&A...631A..77A}. We use the formalism of the shellular rotation hypothesis developed by \citet{1992A&A...265..115Z}, \citet{1998A&A...334.1000M}, and \citet{2004A&A...425..229M} to describe the transport of angular momentum and chemicals by meridional circulation and turbulent shear (vertical and horizontal). 
The transport of angular momentum obeys the advection-diffusion equation
\begin{equation}
    \rho \frac{d}{dt}(r^2 \Omega) = \frac{1}{5 r^2} \frac{\partial}{\partial r} (\rho r^4 \Omega U_2) + \frac{1}{r^2} \frac{\partial}{\partial r}\left(\nu_v r^4 \frac{\partial \Omega}{\partial r}\right),
    \label{eqrot}
\end{equation}
where $\rho$, $r$, $\Omega$, $U_2$, and $\nu_v$ are the density,   radius,   angular velocity,   meridional circulation velocity, and   vertical shellular component of the turbulent viscosity, respectively.\\
Meridional circulation appears through its velocity $U_2$ in Eq.~(\ref{eqrot}), and can be described as a diffusion coefficient $D_{\rm{eff}}$ in Eq.~(\ref{eqdiffu}) for the transport of chemicals, as shown by \citet{1992A&A...253..173C} when assuming a strong turbulent transport in the horizontal direction. The turbulent shear in the vertical and horizontal directions appears as a viscosity $\nu_v$ ($\nu_h$) in Eq.~(\ref{eqrot}) and as a diffusivity $D_v$ ($D_h$) in Eq.~(\ref{eqdiffu}), that is assumed to be proportional to the corresponding viscosity, with a proportionality factor of 1 as usually assumed in stellar evolution models \citep{1992A&A...265..115Z,2018A&A...620A..22M,2008Ap&SS.316...43E,2012A&A...537A.146E}. 
 Several prescriptions exist for both ($D_v, \nu_v$) and ($D_h, \nu_h$), and  stellar evolution models  computed with different combinations of these prescriptions can be found in the literature. As demonstrated by \citet{2013LNP...865....3M} and \citet{2016A&A...587A.105A}, this choice strongly affects the outcome of the models. To explore this aspect we compute models with three different combinations listed in Table~\ref{tab:allparam} and referred to as R1, R2, and R3. We  used these combinations in our previous works  \cite[e.g.][]{2005A&A...440..981T,2016A&A...587A.105A,2018A&A...620A..22M}, motivated by the outcomes of numerical simulations \citep[e.g.][]{2013A&A...551L...3P,2014A&A...566A.110P,2016A&A...592A..59P,2017ApJ...837..133G,2018ApJ...862...36G}. The nomenclature of the models and the detailed expressions of the different turbulent diffusion coefficients are given in Appendices \ref{AnnexeA} and \ref{AnnexeB}, respectively.

\subsubsection{Magnetic braking and initial rotation velocities}
\label{subsub:magneticbraking}
The extraction of angular momentum at the stellar surface due to magnetised winds is accounted for following the formalism by \citet{2015ApJ...799L..23M}.
We use the prescription as written in Eqs.~(7) - (9) in \citet{2019A&A...631A..77A} with
the following values for the parameters m = 0.22 and p = 2.1, which  refer respectively to an exponent related to the magnetic field geometry and the exponent relating rotation and activity. We take $\chi = 14 \equiv \frac{Ro_{\odot}}{Ro_{sat}}$, the ratio of the solar Rossby number to the saturation value of the Rossby number\footnote{The Rossby number is defined here according to \citet{2015ApJ...799L..23M} as $Ro=(\Omega\,\tau_{cz})^{-1}$, where $\tau_{cz}$ is the convective turnover timescale, characterised by the size of the studied convective region divided by the convective velocity.}, at which the magnetic activity indicators saturate. They are fitted on the clusters of different ages to reproduce the rotation velocity dispersion according to \citet{2019A&A...631A..77A}. 
Finally, a last complement parameter linked to magnetised wind braking, $K$, is calibrated so as to reproduce the solar surface rotation at the age of the Sun. We use a value of $K = 7.5\times10^{30}$erg for our models unless otherwise indicated (see also Appendix A).\\

The models with rotation are computed 
as in \citet{2019A&A...631A..77A} for three values of the initial rotation period on the PMS: 1.6, 4.5, and 9.0 days, which will be referred to  as the fast ($^{\textbf{F}}R$), median ($R$), and slow ($^{\textbf{S}}R$) rotating models, respectively. The disc coupling timescale is set at $\tau_{disc}$ = 2.5 Myr for the fast rotators and at $\tau_{disc}$ = 5 Myr for the median and the slow rotators. These values are chosen in agreement with \citet{2015A&A...577A..98G} in order to reproduce the observed rotation spread of open clusters stars.
 
\subsection{Overshooting and penetrative convection} 
\label{sub:penetrativeconvection}

Using the MLT formalism associated with the Schwarzschild criterion for the instability to describe the extent of the convective regions, which is a classical approach in stellar evolution codes, is known to be flawed as the convective edges are defined according to null acceleration instead of null velocity of convective eddies. Convection actually penetrates in the sub-adiabatic layers below (for the convective envelopes) the superadiabatic unstable region, which generates mixing beyond the convective region down to where convective eddies are braked or eroded \citep[e.g.][]{1991A&A...252..179Z}. 
Several formalisms for penetrative convection exist; we test 
three recent ones with an associated turbulent diffusion coefficient that scales with depth.  
We assume that the transport of angular momentum is not impacted by this process.\\
The diffusion coefficients given below enter the expression of $D$ in Eq.~(\ref{eqdiffu}).

\subsubsection{\citet{2017ApJ...845L...6B}}

The formalism proposed by \citet{2017ApJ...845L...6B} is based on 2D and 3D  hydrodynamic simulations of a young Sun on the PMS at 1 Myr and at solar metallicity \citep[][]{2017A&A...604A.125P}. 
 These simulations allow them to characterise the depth of the penetrative convection below the convection zone; they show the existence of extreme events (deep penetrating plumes), which can have an outsized impact on transport mechanisms, especially ones that are meant to be inviscid. The diffusion coefficient obtained by \cite{2017A&A...604A.125P} and reproduced here in Eq.~(\ref{eq:dbaraffe}), describes the mixing in the penetration layers and is characterised by the cumulative distribution function of the maximum penetration depth
obtained in the simulations:  
\begin{equation}
    D_{B}(r) = D_{0} \left[1-\rm{exp}\left(-\rm{exp}\left(-\frac{\frac{r_{bcz}-r}{R}-\mu}{\lambda}\right)\right)\right]. 
    \label{eq:dbaraffe}
\end{equation}
Here $D_0 = (\upsilon_{\rm conv} \times H_p \times \alpha_{\rm MLT})/3$ is the convective turbulent diffusivity (with $\upsilon_{\rm conv}$ the mean velocity of the convective elements obtained from MLT and $\alpha_{\rm MLT}$ the mixing length parameter),  $r$ is the local radius,  $r_{bcz}$ is the radius at the base of the convective zone,  and R is the total radius of the star. The coefficients $\lambda = 6\times10^{-3}$ and $\mu = 5\times10^{-3}$ are as prescribed by \citet{2017ApJ...845L...6B} and obtained from the simulations of \citet{2017A&A...604A.125P}. They are assumed to be independent of the stellar structure and age. All physical quantities are in cgs units (here and throughout the paper). 
The penetration depth of the overshooting is limited by the free parameter $d_{ov}$ linked to the pressure scale-height and adjusted to take into account the limiting effect of stellar rotation.
\citet{2017ApJ...845L...6B} determined that this parameter should be $d_{ov} \approx 0.30 H_p-0.35 H_p$ to reproduce the solar lithium abundance.

\subsubsection{\citet{2019ApJ...874...83A}}

We tested for the first time in a stellar evolution code including rotation the description of penetrative convection by \citet{2019ApJ...874...83A}, which is based on a new model of rotating convection in stellar interiors. Contrary to Eq.~(\ref{eq:dbaraffe}), the penetration depth is now based on the one obtained by \citet{1991A&A...252..179Z}. In his work \citet{1991A&A...252..179Z} linearised the equations of motion in the region of penetration, and the depth of penetration is then solved for given the velocity at the upper boundary of that region, which is assumed to reside in the convection zone. In \citet{2019ApJ...874...83A} the impact of rotation on the convection is accounted for by using a modal convection model for rotating Rayleigh-Benard convection \citep{2014ApJ...791...13B} where it is assumed to be locally valid in the region of penetration. This model has the asymptotic property that the velocity scales as $(\rm{v/v_0}) \propto \rm{Ro}^{1/5}$, with Ro the Rossby number, which is inversely proportional to the angular velocity in the convective region. The depth of the overshooting zone then depends on the pressure scale-height, the convective Rossby number (hence angular velocity), and the thermal diffusivity, and is dynamically estimated. Using this model for penetrative convection and the functional form proposed by \cite{2017A&A...604A.125P}, 
\citet{2019ApJ...874...83A} derive a new expression for the diffusion coefficient (their Eq.~(70)), an approximation of which we use in this paper: 
\begin{equation}
    D_{A}(r) \approx D_0 \left[1-\exp\left(-\exp\left(\frac{r-r_{bcz}}{d_{ov} \times \left(\rm{\frac{v}{v_0}}\right)^{3/2}} + \frac{\mu}{\lambda} \right)\right)\right]. 
    \label{eq:dkyle}
\end{equation}
Here $D_0$, $\mu$, and $\lambda$ are the same as in Eq.~(\ref{eq:dbaraffe}); $d_{ov}$ is the free parameter for controlling the depth of the overshoot; and $\rm{(v/v_0)}$ is the ratio of the velocity of the convective elements when taking rotation into account to the non-rotating inviscid value.  
The scaling between the velocity scales and the angular velocity implies that the diffusion coefficient in Eq.~(\ref{eq:dkyle}) is smaller when the star rotates faster, mimicking the fact that fast rotation inhibits convective motions to penetrate deep into the stably stratified region below. Thus, this model is a combination of the \citet{2017ApJ...845L...6B} fit to their numerical stellar convective penetration simulations and those theoretical results of \citet{2019ApJ...874...83A}. 

\subsubsection{\citet{2019MNRAS.484.1220K}}

\citet{2019MNRAS.484.1220K} propose a diffusive prescription suitable for 1D stellar evolution codes based on hydrodynamical simulations of penetrative convection and overshooting
in a non-rotating Boussinesq spherical shell. In comparison to the simulations of \citet{2017A&A...604A.125P}, these simulations  include   an explicit diffusion, being the Navier-Stokes equations rather than an approximation of the Euler equations. The prescription is given by their Eq. (45), 
\begin{equation}
    D_K (r) = D_0\,\rm{exp} \left(- \frac{(r-r_{bcz})^2}{2 \delta_G^2}\right),
    \label{eq:dKorrea}
\end{equation}
where $D_0$ (denoted $D_{\rm{cz}}$ in the original paper) is the same as in Eq.~(\ref{eq:dbaraffe}) and $\delta_G$ controls the depth of the penetration and writes 
\begin{equation}
    \delta_G \approx 1.2 \left(\frac{E_0 P r}{S Ra_0}\right)^{1/2},
\end{equation}
where $E_0$ is the energy in the non-rotating convection zone, $Pr = \nu / \kappa$ is the Prandtl number defined as the ratio of the viscosity $\nu$ to the thermal diffusivity $\kappa$, S is the stiffness that measures the stability of the interface between the radiative zone and the convective zone, and $Ra_0$ is the Rayleigh number defined by Eq. (11) in \citet{2019MNRAS.484.1220K} as 
\begin{equation}
    Ra_0 = \frac{\alpha_{\rm{th}} g  \left|\frac{dT_0}{dr}-\frac{dT_{\rm{ad}}}{dr}\right| r_0^4}{\kappa \nu},
\end{equation}
with $\alpha_{\rm{th}}$ the thermal expansion coefficient, g the gravity, $\frac{dT_{\rm{ad}}}{dr}$ the adiabatic temperature gradient, $\frac{dT_0}{dr} = \frac{dT_{\rm{ad}}}{dr}|_{r=r_0}$, $r_0$ the outer radius of the convection zone, $\kappa$ the thermal diffusivity, and $\nu$ the viscosity.
\newline
\\ In this work we used Eq.~(\ref{eq:dKorre}) (instead of Eq.~(\ref{eq:dKorrea})), which is a result of fits to the numerical penetrative convection simulations carried out in \citet{2019MNRAS.484.1220K} where the overshooting length is adapted to contain information about the local rotation rate through the convective model of \cite{2019ApJ...874...83A} and its implications for a linearised convective penetration model 
\begin{equation}
    D_K (r) \approx D_0 \; \exp \left(-\frac{(r-r_{bcz})^2}{d_{ov}^2 \times \left(\rm{\frac{v}{v_0}}\right)^2}\right) ,
    \label{eq:dKorre}
\end{equation}
 where $\delta_G$ is approximated using the same principle as for $D_A(r)$ and adding the velocity dependence ($\rm v/v_0$) \citep[][]{2020svos.conf..311A} 

\begin{equation}
    \delta_G \approx d_{ov} \times \rm{\frac{v}{v_0}},
\end{equation}

Due to this inverse dependence on rotation, $D_K(r)$ is also smaller when the star rotates faster. Regarding the initial numerical simulation of \citet{2019MNRAS.484.1220K}, the way that we adapted the rotational dependence with the help of the \citet{2019ApJ...874...83A} should be taken with precautions. However, in the framework of slow-rotators, it is a relevant assumption.

\subsection{Additional transports of chemicals}
\label{sub:addchemicaltransp}

\subsubsection{Tachocline turbulence}
\label{tachocline-formalism}
The tachocline is a shear layer located at the base of the solar convection zone where the radial rotation profile goes from differential in the convective envelope to flat in the radiative interior \citep{1988ESASP.286..149C}. First modelled by \citet{1992A&A...265..106S} in a hydrodynamical framework, it is considered to be the seat of strong turbulence, with an associated diffusion coefficient that can be parametrised with respect to the Brunt-V\"ais\"al\"a frequency, the thickness of the tachocline, and the horizontal turbulent viscosity within it.
Here we present models including the time-dependent expression given by Eq. (15) and (16) in \citet{1999ApJ...525.1032B},
\begin{equation}
    D_{Tach}(\zeta) =  \frac{1}{180}\frac{1}{4}\left(\frac{8}{3}\right)^2 \nu_H \left(\frac{d}{r_{bcz}}\right)^2  \mu^6_4  Q^2_4 \exp(-2\zeta)\cos^2(\zeta), 
    \label{eq:DtachBrun}
\end{equation}
where $\nu_H$ is the horizontal turbulent viscosity; $r_{bcz}$ is the radius at the base of the convective envelope; $\zeta = \mu_4 (r_{bcz}-r)/d$ is a non-dimensional depth; $\mu_4 = 4.933$; $Q_4 \approx \hat{\Omega}/\Omega$, with $\hat{\Omega}=d\Omega(r,\theta)/d \theta$ the latitudinal differential rotation  at the base of the convective envelope; and $d$ is a measure of the tachocline thickness $h \approx d/2$: 
\begin{equation}    
    d(t) = r_{bcz} \left(\frac{2\Omega}{N}\right)^{1/2}  \left(\frac{4K_T}{\nu_H}\right)^{1/4}.
    \label{eq:dtachot}
\end{equation}
The depth of the convective envelope $r_{\rm{bcz}}$, the angular velocity $\Omega$, the horizontal viscosity $\nu_H$, the thermal diffusivity $K_T$, and the Brunt-V\"ais\"al\"a frequency $N$ all vary in time, as predicted by the structure and rotation equations.\\
Defining
\begin{equation}
    C = \frac{1}{180}\frac{1}{4}\left(\frac{8}{3}\right)^2\mu^6_4\exp(-2\zeta)\cos^2(\zeta),
    \label{eq:C}
\end{equation}
we can compute the fully time-dependent equation of $D_{\rm{Tach}}$ as
\begin{equation}
    D_{\rm{Tach}}(t) = C \times \nu_H \left(\frac{d}{r_{bcz}}\right)^2 \left(\frac{\hat{\Omega}}{\Omega}\right)^2 \propto \Omega \nu_h^{1/2}\left(\frac{\hat{\Omega}}{\Omega}\right)^2.
    \label{eq:tachot}
\end{equation}
 The treatment of the meridional circulation in the framework of Zahn's theory is based on the expansion of all the physical quantities, including meridional circulation, in Legendre polynomials. The meridional circulation velocity is expanded to the second-order Legendre polynomials in the original works by \cite{1992A&A...265..115Z} and \cite{1998A&A...334.1000M}, which is the formalism adopted in STAREVOL. In that case, as shown by \cite{2004A&A...425..229M}, the differential rotation in latitude is not explicitly accounted for. An expansion to the fourth-order of the departures from spherical symmetry is required to  simultaneously treat the bulk of a radiative region and its tachocline, which is beyond the scope of this study.  
 Hence, we need a prescription to evaluate  $\hat{\Omega}/{\Omega}$,  and we adopt the same proportionality as in \citet{1999ApJ...525.1032B}, namely $\hat{\Omega} \propto \Omega^{0.7 \pm 0.1}$, which in turn comes from the paper from \citet{1996ApJ...466..384D}, and has been confirmed since then \citep[e.g.][]{2009ASPC..416..375S,2017ApJ...836..192B}, even if some uncertainties subsist \citep[e.g.][who found that it would scale inversely with $\Omega$ in the case of F-type stars]{2012ApJ...756..169A}.\\
 Eq.~(\ref{eq:tachot}) can thus be recast as
 \begin{equation}
    D_{\rm{Tach}}(t) = C \times 0.02 \left(\frac{\Omega^{0.4}}{N^{1/2}}\right) \left(4\nu_H K_T \right)^{1/2},
    \label{eq:tachot1}
\end{equation}
 where we make explicit the proportionality coefficient adopted and the actual expression used in our computations for   $\hat{\Omega}/\Omega$.
 
\subsubsection{Parametric turbulent transport coefficients} 
\label{adhoc-prescriptions}
The physical turbulent processes generating chemical mixing in radiative interiors, and more specifically in the radiative regions bordering convective ones, cannot all be accounted for given our current state of knowledge. Instead, these processes are   parametrised to simulate diffusive turbulent mixing. Specifically, we follow \citet{2000ApJ...529..338R} and \citet{2005ApJ...619..538R} who discussed the use of additional turbulence competing with atomic diffusion to account respectively for observed abundance anomalies in Am and Fm stars and for the lithium abundances of Population II halo low-mass stars. In both classes of objects, atomic diffusion can be very efficient, and comparing models with observations calls for additional transport processes to be able to counteract its effects, as already anticipated in different kinds of stars by a vast amount of  literature \citep[e.g.][]{1929MNRAS..90...54E,2013EAS....63..233V,2015ads..book.....M}. \citet{2000ApJ...529..338R} and \citet{2005ApJ...619..538R}
propose a purely parametric approach to model turbulence, with no assumptions on the underlying physical mechanism (see also \citealt{2006ApJ...645..634T} for a comparison with rotation-induced mixing). Their diffusion coefficient is proportional to $\rho^{-3}$ (see \citealt{1991ApJ...380..238P}) and attached to a specific temperature or to the density at the base of the convective envelope.  
These fixed points control where the turbulence is generated and thus the mixing depth. 
The parametric diffusivities $D_{\rm{T_0}}$ and $D_{\rm{PMa_0}}$ correspond to equations (2) and (3) of \citet{2005ApJ...619..538R}, respectively, 
\begin{equation}
    D_{\rm{T_0}} = 400 D_{\rm{He}}(T_0)\left[\frac{\rho(T_0)}{\rho}\right]^3,
\label{eq:dturb1}
\end{equation}
where $T_0$ is a free parameter corresponding to the temperature at which the diffusivity is set to be 400 times larger than the atomic diffusion coefficient for He (e.g.  $D_{\rm{He}}(T_0))$, which is computed with the analytical approximation, as advised by \citet{2000ApJ...529..338R}, and $\rho(T_0)$ is the density at the location where $T \equiv T_0$:
\begin{equation}
    D_{\rm{PMa_0}} = a_0 \left[\frac{\rho_{\rm{bcz}}}{\rho}\right]^3
\label{eq:dturb2}
\end{equation}
Here $a_0$ is a free factor and $\rho_{\rm{bcz}}$ is the density at the base of the convective zone. To avoid numerical issues a minimum value of 1 is imposed on $D_{\rm{T_0}}$ and $D_{\rm{PMa_0}}$.
\begin{sidewaystable*}[p]
    \centering
    \caption{Parameters of the rotating models (column 1) related to the prescriptions for horizontal and vertical turbulent viscosities (columns 2 and 3 respectively), the formalism for the overshoot (column 4), the value adopted for the free parameter $d_{ov}$ that controls the depth of the overshooting (column 5), the wind torque K (column 6), the value of the additional constant viscosity $\rm \nu_{add}$ (column 7), the parameters describing its evolution ($\nu_0$ and $\alpha$) according to Eq~\ref{eqnuaddt} (column 8), and the adopted turbulence coefficient (column 9).}
    \begin{tabular}{ccccccccc}
    \hline \hline
    Model & $D_h$ & $D_v$ & Overshoot & $d_{ov}$ & K (erg) & $\nu_{\rm add}$ ($\rm{cm^2.s^{-1}}$) & $\nu_0,\alpha$ & Turbulence \\ 
    \hline
    $_{\rm solid}R_A$ & - & - & $D_A$ & 0.0325 & $1.1 \times 10^{31}$  & - & - & - \\
    R1 & \text{\citet{2018A&A...620A..22M}} & \text{\citet{1992A&A...265..115Z}} & - & - & $7.5 \times 10^{30}$ & - & - & -  \\
    $R1_{B.S}$ & \text{\citet{2018A&A...620A..22M}} & \text{\citet{1992A&A...265..115Z}} & $D_B$ & 0.340 & $7.5 \times 10^{30}$ & - & - & - \\
    $R1_{B.E}$ & \text{\citet{2018A&A...620A..22M}} & \text{\citet{1992A&A...265..115Z}} & $D_B$ & 0.100 & $7.5 \times 10^{30}$ & - & - & - \\
    $R1_A$ & \text{\citet{2018A&A...620A..22M}} & \text{\citet{1992A&A...265..115Z}} & $D_A$ & 0.0325 & $7.5 \times 10^{30}$ & - & - & -  \\
    $R1_K$ & \text{\citet{2018A&A...620A..22M}} & \text{\citet{1992A&A...265..115Z}} & $D_K$ & 0.055 & $7.5 \times 10^{30}$ & - & - & -  \\
    $_{\nu}R1_A$ & \text{\citet{2018A&A...620A..22M}} & \text{\citet{1992A&A...265..115Z}} & $D_A$ & 0.0325 & $7.5 \times 10^{30}$ & $3.5 \times 10^{4}$ & - & -  \\
    $_{\rm \nu.spada}R1_A$ & \text{\citet{2018A&A...620A..22M}} & \text{\citet{1992A&A...265..115Z}} & $D_A$ & 0.0325 & $1.2 \times 10^{30}$ & $2.5 \times 10^{5}$ & - & -  \\
    $_{\rm \nu.spada(t)}R1_A$ & \text{\citet{2018A&A...620A..22M}} & \text{\citet{1992A&A...265..115Z}} & $D_A$ & 0.0325 & $7.5 \times 10^{30}$ & - & $2.5 \times 10^{4}$, 0.5 & -  \\
    $_{\rm \nu2.spada(t)}R1_A$ & \text{\citet{2018A&A...620A..22M}} & \text{\citet{1992A&A...265..115Z}} & $D_A$ & 0.0325 & $7.5 \times 10^{30}$ & - & 100, 12 & -  \\
    $_{\nu}R1_A^{T6.425}$ & \text{\citet{2018A&A...620A..22M}} &  \text{\citet{1992A&A...265..115Z}} & $D_A$ & 0.0325 & $7.5 \times 10^{30}$ & $3.5 \times 10^{4}$ & - & $\rm D_{T6.425}$  \\
    $_{\nu}R1_A^{T6.42}$ & \text{\citet{2018A&A...620A..22M}} & \text{\citet{1992A&A...265..115Z}} & $D_A$ & 0.0325 & $7.5 \times 10^{30}$ & $3.5 \times 10^{4}$ & - & $\rm D_{T6.42}$  \\
    $_{\nu}R1_A^{PM5000}$ & \text{\citet{2018A&A...620A..22M}} & \text{\citet{1992A&A...265..115Z}} & $D_A$ & 0.0325 & $7.5 \times 10^{30}$ & $3.5 \times 10^{4}$ & - & $\rm D_{PM5000}$  \\
    $_{\nu}R1_A^{Tach}$ & \text{\citet{2018A&A...620A..22M}} & \text{\citet{1992A&A...265..115Z}} & $D_A$ & 0.0325 & $7.5 \times 10^{30}$ & $3.5 \times 10^{4}$ & - & $\rm D_{Tach}$ \\
    $R2_A$ & \text{\citet{1992A&A...265..115Z}} & \text{\citet{1997A&A...317..749T}} & $D_A$ & 0.0325 & $7.5 \times 10^{30}$ & - & - & - \\
    $R2^{'}_A$ & \text{\citet{1992A&A...265..115Z}} & \text{\citet{1997A&A...317..749T}} & $D_A$ & 0.0325 & $3.0 \times 10^{30}$ & - & - & -  \\
    $_{\nu}R2^{''}_A$ & \text{\citet{1992A&A...265..115Z}} & \text{\citet{1997A&A...317..749T}} & $D_A$ & 0.0325 & $4.5 \times 10^{30}$ & $2.5 \times 10^{4}$ & - & -  \\
    $R3_A$ & \text{\citet{2004A&A...425..243M}} & \text{\citet{1992A&A...265..115Z}} & $D_A$ & 0.0325 & $7.5 \times 10^{30}$ & - & - & -  \\
    $_{\nu}R3_A$ & \text{\citet{2004A&A...425..243M}} & \text{\citet{1992A&A...265..115Z}} & $D_A$ & 0.0325 & $7.5 \times 10^{30}$ & $3.5 \times 10^{4}$ & - & - \\
    \hline
    \end{tabular}
    \tablefoot{References. $D_B$: \citet{2017ApJ...845L...6B}, $D_A$: \citet{2019ApJ...874...83A}, $D_K$: \citet{2019MNRAS.484.1220K}, $\rm D_{T.425}$: \citet{2005ApJ...619..538R}, $\rm D_{T.42}$: \citet{2005ApJ...619..538R}, $\rm D_{PM5000}$: \citet{2005ApJ...619..538R}, $\rm D_{Tach}$: \citet{1999ApJ...525.1032B}}
    \label{tab:allparam}
\end{sidewaystable*}

\section{Type I rotating models with atomic diffusion and penetrative convection}
\label{MODELS}
As discussed in the introduction, classical models do not account for the evolution of Li with time observed in solar-type stars. However, we computed  such a model for comparison purposes (C models, see Table~\ref{table_calib}), but we focus our discussion on models including rotation (R models), which all include atomic diffusion. In this section we present  Type I rotating models, where the transport of angular momentum is driven only by meridional circulation and shear turbulence  (\S~\ref{sub:stellarrotation}); we focus on median rotators (see \S~\ref{subsub:magneticbraking}). We  discuss the impact of the initial rotation rate in \S~\ref{sec:improved}.

\subsection{General behaviour}
\label{sub:General}
\begin{figure}[t]
         \center
         \includegraphics [width=90mm]{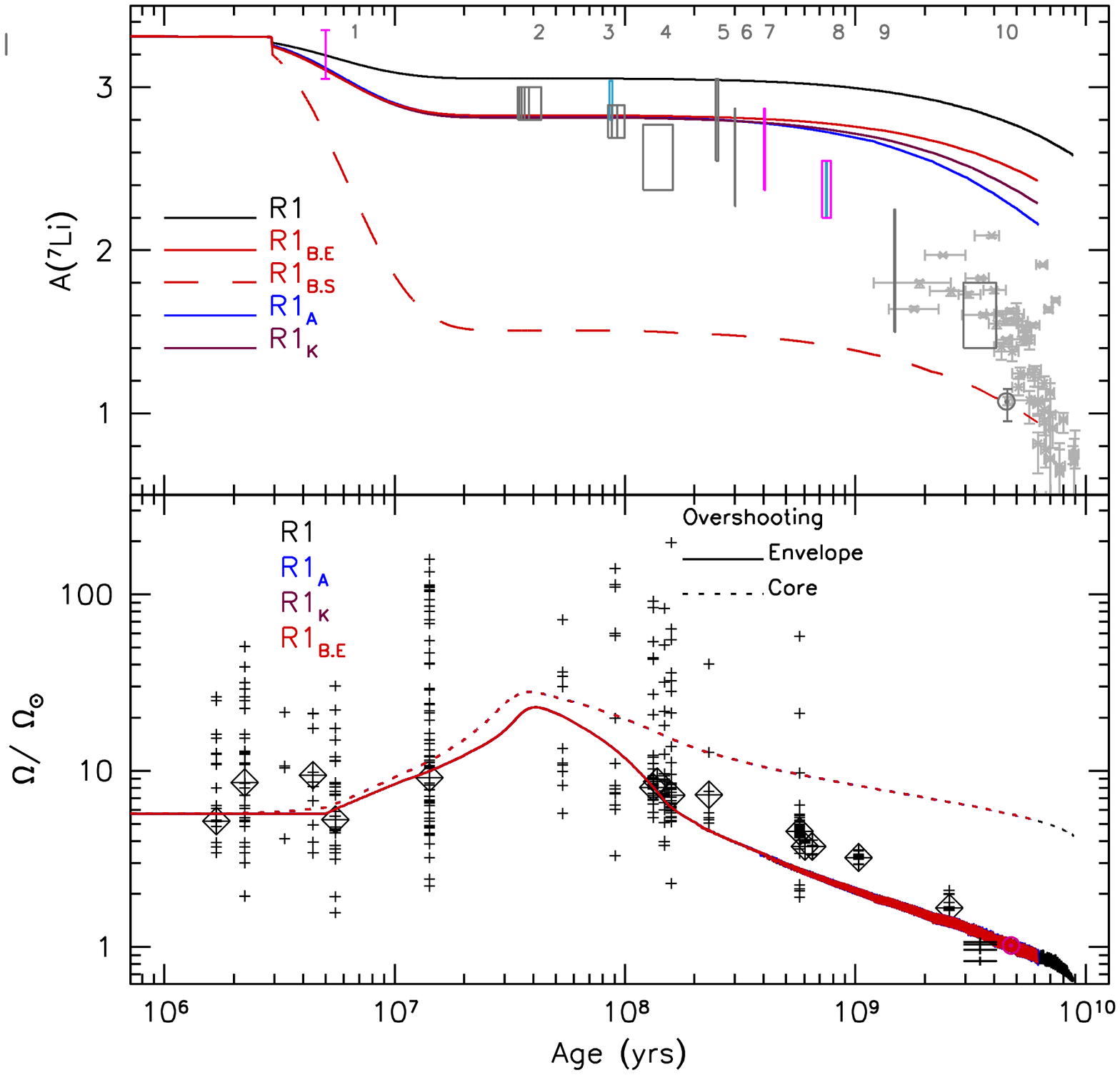}
         \caption{(Top) Lithium surface abundance evolution with time for the $R1$ models including different overshoot prescriptions (colour-coded). Individual points are data for solar twins  \citep{2019MNRAS.485.4052C}. Boxes are for Li observations of solar-type stars in different open clusters (\citealt{2005A&A...442..615S}; \citealt{2020MNRAS.492..245C} for M~67) with ages from \citet{2019A&A...623A.108B}. The numbers 1 to 10 identify the clusters: 1) NGC 2264, 2) IC2391, IC2602 and IC4665, 3) $\alpha$ Per, Pleiades, Blanco I, 4) NGC2516, 5) M34, 6) NGC6475, 7) M35, 8) Praesepe, NGC6633, 9) NGC752, and 10) M67. The colour of the boxes indicates the [Fe/H] value: \citet{2016A&A...585A.150N}: pink: -0.17 to -0.05; grey: -0.05 to 0.05; light blue:  0.05 to 0.16.  
             (Bottom) Evolution of the angular velocity of the convective envelope and of the radiative core (in solar units $\Omega_{\odot} = 2.86\times10^{-6} s^{-1}$; solid and dotted lines, respectively) vs  age. The observational data is from \citet{2015A&A...577A..98G}, except the four stars of M67  from \citet{2016ApJ...823...16B}. Crosses are for individual stars; open diamonds show the $50{th}$ percentiles of the observed rotational distributions in each cluster.}
         \label{fig:evolLi1}
\end{figure}

 We start with model $R1$\footnote{See Appendix~\ref{AnnexeA} for model notation.} for which we adopt the same prescriptions for the horizontal and vertical shear induced turbulent viscosities (\citealt{2018A&A...620A..22M} and \citealt {1992A&A...265..115Z}, respectively; see Table~\ref{tab:allparam}), the same initial rotation period (4.5~days), and the same disc lifetime (5 Myr) as for the 1~M$_\odot$, Z$_\odot$ median rotator model of \citet{2019A&A...631A..77A}. The predicted evolution of the Li surface abundance and of the mean core and envelope angular velocities is shown in Fig.~\ref{fig:evolLi1}. As the envelope is convective and is assumed to rotate as a solid body, we have $\overline{\Omega}_{\rm{conv}}(t) \equiv \Omega_{\rm{surf}}(t)$. The averaged core angular velocity represents the angular velocity of a solid body of equal angular momentum to that of the entire radiation zone, and is defined as in \citet{2019A&A...631A..77A}, 
 \begin{equation}
    \overline{\Omega}_{\rm{rad}} = \frac{\int\limits_{0}^{M_{\rm{BCE}}} r^2 \Omega dm}{\int\limits_{0}^{M_{\rm{BCE}}} r^2 dm},
    \label{eq:omegarad}
 \end{equation} 
 with $M_{\rm{BCE}}$ the mass coordinate at the base of the convective envelope.
 
 As discussed in \citet{2019A&A...631A..77A}, such a model  reproduces well the 50${th}$ percentiles of the period distributions of most of the clusters for solar-type stars in open clusters from the early PMS to the age of the Sun and beyond. This behaviour is mainly driven by the extraction of angular momentum at the surface via magnetised winds modelled following \citet{2015ApJ...799L..23M}.
 As seen in Fig.~\ref{fig:evolLi1}, we confirm that the meridional circulation and shear turbulence are too weak to enforce the coupling between the core and the surface, leading to a core spinning too fast at the age of the Sun compared to the internal rotation profile inferred through helioseismology (see  \S~\ref{subsub:dhdv}).
 On the other hand, the associated transport of chemicals is dominated by vertical shear-induced turbulence, which remains weak during the entire evolution (see \S~\ref{sub:optimalprecript}, and   \citealt{2018A&A...620A..22M}) and only partially counteracts atomic diffusion. Consequently, at the age of the Sun the predicted lithium abundance is $\sim$1.5~dex higher than observed.
\subsection{Penetrative convection}
\label{penetrativeconvection}
As extensively discussed in the literature, model predictions for PMS Li depletion strongly depend on the treatment of convection and of penetrative convection \cite[see e.g.][]{1994ApJS...90..467D,2002ApJ...566..419P,2017ApJ...845L...6B,2017A&A...598A..64T}. Here we test three different prescriptions for penetrative convection beyond the convective envelope that depend on the evolution of the internal angular velocity profile (\citealt{2017ApJ...845L...6B}, \citealt{2019ApJ...874...83A}, and \citealt{2019MNRAS.484.1220K}; see \S~\ref{sub:penetrativeconvection} and Table~\ref{tab:allparam}).
Penetrative convection is assumed to only  transport chemicals. It does not affect the surface and internal angular velocity evolution, which behave as in $R1$ (Fig.~\ref{fig:evolLi1}). 

Concerning the impact on Li, and as already shown by \citet{2017ApJ...845L...6B}, its evolution when assuming a constant value for the extent of the overshoot region $d_{ov}$ does not match the observed Li behaviour with time. As shown in Fig.~\ref{fig:evolLi1}, the models $R1_{B.S}$ and $R1_{B.E}$ implementing Eq.~\ref{eq:dbaraffe} with $d_{ov} = 0.34 H_p $ and $0.1 H_p$, respectively, fit either the solar Li abundance or the Li abundances in the youngest open clusters (between 10 Myr and 100 Myr). \citet{2017ApJ...845L...6B} thus proposed to vary the depth of the overshooting zone depending on rotation, as supported by numerical studies \citep{2003A&A...401..433Z,2007IAUS..239..417B,2017ApJ...836..192B}.
They reproduce   the Li temporal evolution fairly well when adopting $d_{ov} = 0.1 H_p$ when the star is rapidly rotating ($\Omega > 5 \Omega_{\odot}$, i.e. typically on the late PMS and around the ZAMS, and similarly to what we find) and a much deeper overshooting zone ($d_{ov} = 1 H_p$) for slower, more evolved rotating stars. 

We decided  not to fine-tune the depth of the overshooting zone for the slow rotators (after $\sim$ 1 Gyr), however, considering that 
other slow mixing processes may be responsible for Li depletion during this phase (see \S~\ref{sub:addchemicaltransp2}). 
We found the same behaviour as above with the prescription by \citet{2019ApJ...874...83A} 
that accounts for the impact of the rotation on penetrative convection efficiency, and with the prescription by \citet{2019MNRAS.484.1220K}. In both cases (models $R1_{A}$ and $R1_{K}$, Fig.~\ref{fig:evolLi1}, Table~\ref{tab:allparam}) we adjusted the free parameter $d_{ov}$ to reproduce the Li behaviour observed in the youngest open clusters around 30 Myr as in model $R1_{B.E.}$.\\ The Li evolution in all these models is characterised by a mild decrease from A($^7$Li) = 3.3 dex to A($^7$Li) = 2.8 dex at $\sim$ 15 Myr, followed by a plateau and a further slow decrease later on the MS. The convective envelope slowly recedes along the MS, which increases the term $(r-r_{bcz})$ in the exponential of Eqs.~(\ref{eq:dbaraffe}), (\ref{eq:dkyle}), and (\ref{eq:dKorre}), leading to an increased efficiency of the slow mixing in the overshooting region that slowly depletes Li at the surface. In models $R1_A$ and $R1_K$ the transport in the overshoot region also depends on the rotation rate, which affects the depth of convective penetration via the ratio $(\rm{v/v_0})$. This results in a multiplying factor to the convective turbulent diffusivity $D_0$ in Eq.~(\ref{eq:dkyle}) larger than that in Eq.~(\ref{eq:dKorre}), hence a more efficient mixing and a greater Li depletion in model $R1_A$ than in model $R1_K$.

\subsection{Impact of turbulent diffusion modelling}
\label{subsub:dhdv}

\begin{figure}[t]
         \center
         \includegraphics [width=90mm]{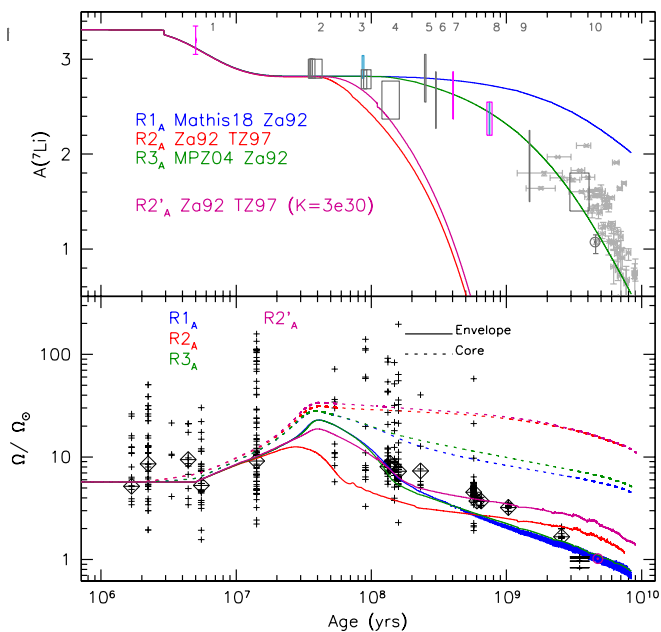}
         \caption{Same as Fig.~\ref{fig:evolLi1}, but for the $R1_A$, $R2_A$, $R2'_A$, and $R3_A$ models (details in Table~\ref{tab:allparam}).}
         \label{fig:evolLi2}
 \end{figure}

From now on we consider models for median rotators including atomic diffusion, penetrative convection according to \citet{2019ApJ...874...83A} with the calibration described above, and rotation-induced transport of chemicals and angular momentum associated with meridional circulation and shear turbulence.
We explore the impact of different sets of prescriptions for the horizontal and vertical shear diffusivities ($D_v$ and $D_h$, respectively)  reported in Table~\ref{tab:allparam}. 
Figure~\ref{fig:evolLi2} shows the evolution of $A(^7Li)$ and core and envelope angular velocities as a function of age for the corresponding models $R1_A$ (\S~\ref{penetrativeconvection}), $R2_A$,  $R2^{'}_{A}$, and $R3_A$. Figure~\ref{fig:profomegasun} presents the internal rotation profiles predicted by these models at the age of the Sun.\\

Models $R2_A$ and $R2^{'}_{A}$ include the same prescriptions for $D_v$ and $D_h$ (\citealt{1997A&A...317..749T} and \citealt{1992A&A...265..115Z}, respectively) as in \citet{2005Sci...309.2189C},
and they differ from each other by the adjustment of the torque parameters to better fit the surface rotation rate (Table \ref{tab:allparam}).
In both cases we confirm that these prescriptions fail to reproduce the Li evolution with time and the internal rotation rate of the Sun.
In particular, the (too) strong Li depletion is driven by
 the vertical shear coefficient prescribed by \citet{1997A&A...317..749T}, which is much greater than that of \citet{1992A&A...265..115Z}, and which is fed by the very strong differential rotation inside the star during most of its evolution beyond 20 Myr. 
A similar behaviour for Li was found by \citet{2016A&A...587A.105A} using the same prescription for $D_v$ but that of $D_h$ from \citet{2004A&A...425..243M}, although this combination provides a stronger coupling between the core and the surface (which is still irreconcilable with the solar rotation profile). We thus discard the vertical shear diffusivity as prescribed by \citet{1997A&A...317..749T} based on these results on lithium; however,  this remains  an open question considering that additional transport mechanisms for angular momentum may lead to different conclusions (see e.g. \citealt{2005Sci...309.2189C}, who invoke the effects of internal gravity waves). 

Model $R3_A$ provides a very good fit of the observed evolution of both the surface rotation period and the Li abundance. The rotational evolution of this model is very similar to that of model $R1_A$, as it is dominated by vertical shear in the early phases and by the surface extraction of the angular momentum by the magnetised winds beyond the ZAMS. Stronger Li depletion is achieved, however, thanks to a steeper angular velocity profile below the convective envelope (see Fig.~\ref{fig:profomegasun} for a snapshot at the age of the Sun), which translates into an enhanced turbulent diffusive transport between the base of the convective envelope and the region where Li is destroyed by proton capture. This behaviour is directly related to the modelling of the horizontal shear turbulent viscosity. Accounting for an additional source of horizontal shear turbulence when using the \citet{2018A&A...620A..22M} prescription, as in model $R1_A$, was shown in this previous paper to lead to a lower vertical shear turbulent viscosity than when using the prescription from \citet{2004A&A...425..243M}, as in model $R3_A$. 
In the absence of astero- and helioseismic constraints $R3_A$ would be the best model. However, it also predicts a strong internal differential rotation at the age of the Sun. 

\subsection{Conclusions on Type~I models}
\label{conclusionTypeI}
We confirm that all current prescriptions for shear-induced turbulence fail to reproduce simultaneously the internal rotation profile of the Sun and the surface constraints (see e.g. \citealt{2016A&A...587A.105A,2019A&A...631A..77A} and references therein for similar Type I rotating models without atomic diffusion and penetrative convection; see also
\citealt{2013A&A...555A..54C,2013A&A...549A..74M,2017A&A...599A..18E,2019A&A...626L...1E,2018A&A...620A..22M,2019ARA&A..57...35A,2020arXiv200702585D}).
We favour, however, the prescriptions for $D_v$ and $D_h$ from \citet{2018A&A...620A..22M} and \citet {1992A&A...265..115Z}, respectively. For the modelling of horizontal turbulence, we recall that the treatment of \citet[][]{2018A&A...620A..22M} is the first one that accounts for the action of stratification and rotation on horizontal turbulent motions.
We couple it with the prescription for the vertical turbulent transport, which is the only one that has been validated by direct numerical simulations \citep[][]{2013A&A...551L...3P,2017ApJ...837..133G}. 
Their coupling with angular momentum wind extraction is done with the help of \citet{2015ApJ...799L..23M}; they reproduce well the evolution of the surface rotation rate with time as observed in solar-type stars in open clusters. 
Additionally, the three prescriptions that we tested for rotationally dependent penetrative convection 
help the models more closely fit the Li abundances observed in the youngest clusters (after calibration). Given our choice for the treatment of shear turbulence, accounting for Li in the more advanced stages requires either a very strong dependence between convective penetration depth and rotation, or an additional mixing process that may depend, or not, on the missing transport of angular momentum.

\section{Improving the models}
\label{sec:improved}
In this section we explore  the possibilities to fit all the observational constraints by including additional transport processes for both angular momentum and chemical species. All the models discussed in this section include rotation-induced mixing, atomic diffusion, and penetrative convection.

\subsection{Transport of angular momentum by additional viscosity}
\label{subsub:transangtmomentum} 
Several processes have been proposed to explain the strong coupling between the core and the surface of low-mass stars, and to flatten the angular velocity profile at the solar age and beyond (in particular, internal gravity waves and processes related to magnetism;  \citealt{1993A&A...279..431S,2002A&A...381..923S,2005Sci...309.2189C,2005A&A...440..653M,2005A&A...440L...9E,2010A&A...519A.116E,2010ApJ...716.1269D,2013A&A...554A..40C,2014ApJ...796...17F,2014ApJ...788...93C,2015A&A...573A..80R,2015A&A...579A..31B, 2019A&A...621A..66E,2019A&A...631L...6E,2019MNRAS.485.3661F,2017A&A...605A..31P}). However, no complete solution has been found yet. 
Parametric studies thus remain   necessary to estimate the efficiency of the missing transport processes, and to potentially determine their nature. This is the approach we chose here.

We follow \citet[][see also \citealt{2014sf2a.conf..483L} and \citealt{2016A&A...589A..23S}]{2012A&A...544L...4E,2019A&A...621A..66E} who proposed  introducing a parametric vertical viscosity $\nu_{\rm{add}}$ in the equation describing the transport of angular momentum to reproduce the core rotation rates of SGB and RGB stars. In this context, Eq.~(\ref{eqrot}) becomes 
\begin{equation}
    \rho \frac{d}{dt}(r^2 \Omega) = \frac{1}{5 r^2} \frac{\partial}{\partial r} (\rho r^4 \Omega U_2) + \frac{1}{r^2} \frac{\partial}{\partial r}\left((\nu_{v}+\rm{\nu_{add}}) \textit{r}^4 \frac{\partial \Omega}{\partial r}\right).
\end{equation}

There $\nu_{\rm{add}}$ was assumed to be either constant in time or dependant on the stellar rotation.
Possible variations of $\nu_{\rm{add}}$ within stellar interiors were not considered.
We tested both options and in the second case, we followed \citet[][their Eq.~3]{2016A&A...589A..23S} and assumed that the angular momentum transport efficiency depends on the radial rotational shear, i.e.
\begin{equation}
    \nu_{\rm{add}}(t) = \nu_{0} \times \left(\frac{\overline{\Omega}_{rad}}{\overline{\Omega}_{conv}}\right)^{\alpha}
    \label{eqnuaddt}
,\end{equation}
where $\nu_0$ and $\alpha$ are free parameters, and $\overline{\Omega}_{rad}$ and $\overline{\Omega}_{conv}$ are the mean angular velocity in the radiative interior and convective envelope, respectively, as defined in \S~\ref{sub:General}.
We computed models with the three different combinations of turbulent shear prescriptions discussed in \S~\ref{subsub:dhdv} for different values of $\nu_{\rm{add}}$ and of $\alpha$ and $\nu_0$ (Table~\ref{tab:allparam}).

\begin{figure}[t]
         \center
         \includegraphics [width=90mm]{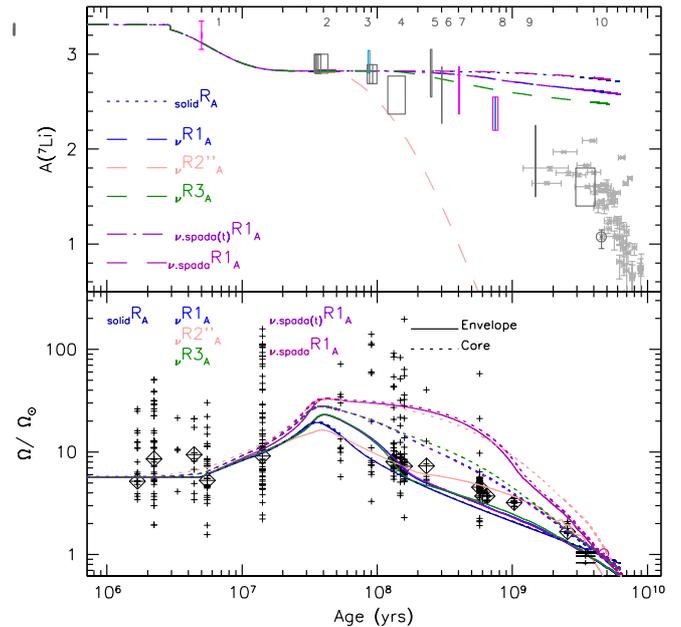}
         \caption{Same as Fig.~\ref{fig:evolLi1}, but for the models with a fixed viscosity $\rm{\nu_{add}}$ ($_{\nu}R1_A$, $_{\nu}R2''_A$, $_{\nu}R3_A$, and $_{\rm{\nu.spada}}R1_A$), a time-dependent viscosity ($_{\rm{\nu.spada(t)}}R1_A$), and a model with enforced solid-body rotation ($_{\rm{solid}}R_A$).
         }
         \label{fig:evolLi3}
 \end{figure}

\begin{figure}[t]
         \center
         \includegraphics [width=90mm]{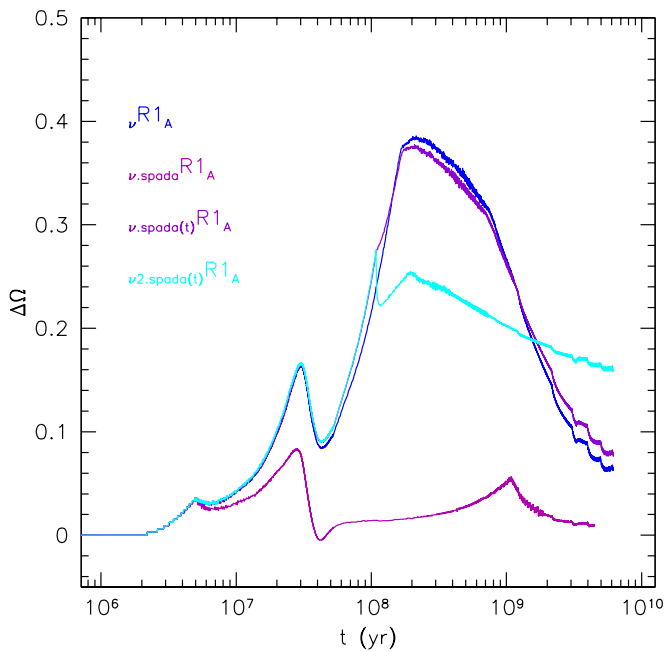}
         \caption{Differential rotation evolution with time for the $_{\nu}R1_A$, $_{\rm{\nu.spada}}R1_A$, $_{\rm{\nu.spada(t)}}R1_A$, and $_{\rm{\nu2.spada(t)}}R1_A$ models (details in Table~\ref{tab:allparam}). 
         }
         \label{fig:evolrotdiff}
 \end{figure}

\begin{figure}[ht]
         \center
         \includegraphics [width=90mm]{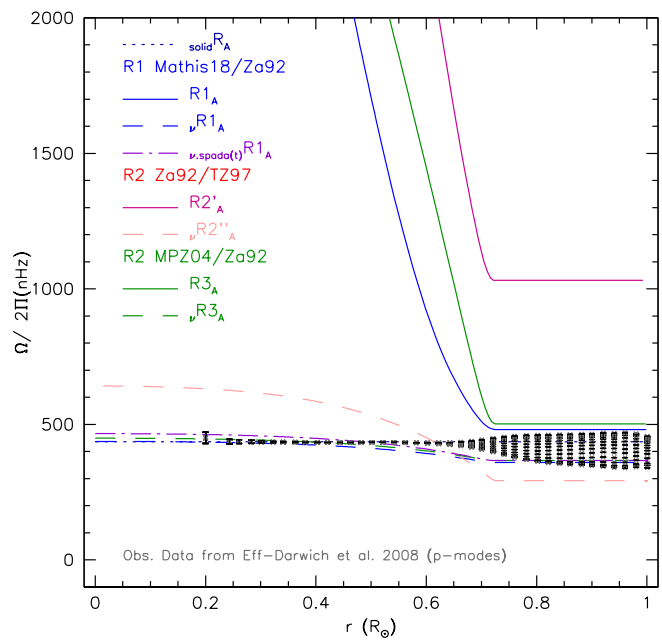}
         \caption{Angular velocity profiles vs the radius at solar age for models presented in Fig. \ref{fig:evolLi3} compared to the rotation profile of the Sun (in dark grey) obtained from helioseismology from \citet{2008ApJ...679.1636E}. The solid line, the dotted line, the dashed line, and the dot-dashed line refer to the differential, solid, $\rm{\nu_{add}}$, and $\nu_{\rm{add}}(t)$ rotation cases, respectively.}
         \label{fig:profomegasun}
 \end{figure}

The impact of the additional viscosity on the internal rotation profile and on the Li depletion and surface rotation rate can be seen in Figs.~\ref{fig:evolLi3}, \ref{fig:evolrotdiff}, and \ref{fig:profomegasun}. In Fig.~\ref{fig:evolrotdiff} we show the level of internal differential rotation
\begin{equation*}
    \Delta \Omega = \frac{\overline{\Omega}_{rad} - \Omega_{surf}}{\overline{\Omega}_{rad} + \Omega_{surf}} 
\end{equation*}
with $\Omega_{surf}$ and $\overline{\Omega}_{rad}$ defined in \S~\ref{sub:General}. 

Clearly, the higher the value of  $\nu_{\rm{add}}$, or the higher the value of $\alpha$ for a given $\nu_0$, the earlier and the stronger the coupling between the core and the surface.
In our models, the values for $\nu_{\rm{add}}$ required to obtain a flat internal rotation profile at the age of the Sun while accounting for the surface rotation constraints vary between 2.5$\times 10^4$ and 4$\times 10^4$ cm$^2$ s$^{-1}$ depending on the adopted prescriptions for $D_v$ and $D_h$. This is in agreement with the values required for the models to fit the asteroseismic data in subgiants and red giant stars ($\approx 10^4$ to $3\times10^4$; \citealt{2012A&A...544L...4E,2016A&A...589A..23S}). In addition, in this case the evolution of the modelled surface rotation rate matches the observed values fairly well all along the evolution of the stars. When we adopt the value of $2.5\times 10^5 \rm cm^2 s^{-1}$ (model $_{\rm{\nu.spada}}R1_A$) advocated by \citet{2016A&A...589A..23S} for their solar benchmark, the internal rotation profile flattens very early in the evolution. In that case, a lower value for the magnetic braking parameter K is needed to fit the solar surface rotation rate\footnote{K = $1.2\times10^{30}$erg (see Table
~\ref{tab:allparam}) to be compared with K = $6.3\times10^{30}$erg advised by \citealt{2019ApJ...870L..27M}.
}, but this destroys the agreement between the model predictions and the observed evolution of the surface rotation rate on the MS (between 100 Myr and 2 Gyr).
When assuming a dependency of the transport of angular momentum with the radial rotational shear (Eq.~\ref{eqnuaddt}), we obtain a reasonable rotation profile at the age of the Sun with $\nu_0 = 2.5\times10^4 \rm{cm^2 s^{-1}}$ and $\alpha = 0.5$ (model $_{\rm{\nu.spada (t)}}R1_A$), where we keep $\nu_0$ of the same order as the constant $\nu_{\rm{add}}$ calibrated above. We also obtain reasonable rotation profiles at solar age with other combinations (e.g. $\nu_0 = 100\,\rm cm^2 s^{-1}$, $\alpha = 12$, model $_{\rm{\nu2.spada(t)}}R1_A$, and $\nu_0 = 1000\,\rm{cm^2 s^{-1}}$, $\alpha = 9$, not shown); in these cases, because of the high value of $\alpha$, the differential rotation is reduced earlier on the MS and slowly continues receding after 100 Myr. It results in a lower surface rotation velocity than for model $_{\rm{\nu.spada (t)}}R1_A$ when $K$ is not re-adjusted. 
\newline
\\In summary, in the absence of asteroseismic data that could reveal the actual internal rotation profile in MS stars younger than the Sun, we cannot better constrain the efficiency of the transport of angular momentum or the nature of the underlying mechanism.
We show, however, that the mean evolution of the surface rotation rate can be fairly well reproduced by the models with moderate values for the additional viscocity and with magnetic braking efficiency in agreement with the calibration value from \citet{2019ApJ...870L..27M}. However, with the stronger viscosity adopted in model $_{\rm{\nu.spada}}R1_A$, no value for the magnetic braking parameter K can reconcile the observed periods along the entire evolution. The same problem occurs for the model where we assume solid-body rotation all along the evolution (models $_{\rm solid}R_A$). This provides a hint that rigid rotation may not be achieved very early on the MS in solar-type stars. This partially agrees with previous works by \citet{2013A&A...556A..36G,2015A&A...577A..98G}, and by \citet{2015A&A...584A..30L} and \citet{2020A&A...636A..76S} for solar-type stars, who find, using two-zone models for the angular momentum transport, that quasi solid-body rotation should be achieved around 1 Gyr in order to fit the same observations. Our models, where the transport of angular momentum is self-consistently treated, resulting in  fully resolved angular velocity radial profiles, seem to indicate that quasi rigid rotation may be reached even later in the evolution (see Fig.~\ref{fig:evolrotdiff}, where $\Delta \Omega = 0$ corresponds to a solid-body rotation and where increasing values of $\Delta \Omega$ correspond to higher differential rotation). 
 
Finally, Li depletion is slightly less efficient in the models with parametric diffusivity than in the corresponding Type~I models, due to the weaker turbulent shear. In particular, model $_{\nu}R3_A$ does not reproduce the Li depletion on the MS contrary to what was obtained for model $R3_A$.
However, Li depletion remains too strong in model $_{\nu}R2''_A$ as it is dominated by the vertical shear associated with the D$_v$ from \citet{1997A&A...317..749T}. 
In \S~\ref{subsub:dhdv} we favoured the prescriptions for $D_v$ and $D_h$ included in models $R1$ and $_{\nu}R1$. We see that model $_{\nu}R1$ reproduces the evolution of surface rotation along time, and predicts an almost flat rotation profile at the age of the Sun. 
However, this model does not reproduce the Li depletion observed in open clusters beyond 500 Myr, in solar twins, and in the Sun. An additional transport of chemicals is consequently needed to increase the mixing and the depletion of the lithium during the MS. 

\subsection{Additional transport for chemicals}
\label{sub:addchemicaltransp2}

 \subsubsection{Tachocline mixing}

As explained in \S~\ref{tachocline-formalism} the tachocline is the transition layer from the latitudinally differentially rotating convective envelope and the uniformly rotating radiative core. It can be the seat of strong anisotropic turbulent transport when assuming that the dynamics of this layer is driven by hydrodynamical mechanisms \citep{1992A&A...265..106S}. We compute a model $_{\rm{\nu}}R1_A^{Tach}$ using   Eqs.~(\ref{eq:dtachot}) and (\ref{eq:tachot1}).  
Model $_{\rm{\nu}}R1_A^{Tach}$ is shown as a medium orange dashed line in Fig. \ref{fig:evolLi4}. This model implements for the first time the self-consistent computation of the tachocline thickness according to \citet{1992A&A...265..106S}. It appears to become thinner as the star evolves from the PMS to the age of the Sun\footnote{Age(yrs)/h $(R_{\odot})$: $10^7/0.05$; $5.10^7/0.065$; $10^9/0.03$; $4.57\,10^9/0.035$.} at which time it extends over 0.035 R$_\odot$. This value is larger, yet compatible with the estimate of $\approx 0.02 R_\odot$ given by \citet{1999ApJ...516..475E} from helioseismology. The efficiency of the transport is also driven by a value of $D_{\rm{Tach}}(t)$ that varies between $\approx 10^9 - 10^{10}$ cm$^2$s$^{-1}$.
At young ages, the transport is very efficient and prevails over the penetrative convection to transport Li because of the deep location of the base of the convective envelope. It leads to an early and strong Li depletion ($\approx 1$ dex), which is too large to reproduce the Li evolution in young open clusters before 1 Gyr. The same result was obtained by \citet{2002ApJ...566..419P} with a parametric treatment of the tachocline depth and turbulent diffusivity as in \citet{1999ApJ...525.1032B}. As the star evolves and the convective envelope becomes shallower, the thickness of the tachocline does not increase enough, and even decreases, so that the tachocline becomes inefficient to transport the Li after 20 Myr whatever the value of $D_{\rm{Tach}}$. This prediction differs from that of \citet{2002ApJ...566..419P}, who achieved a solar Li abundance by the age of the Sun. Such a difference is most probably due to the parametrisation they adopted for the thickness of tachocline.
\newline
\\In its current form and with the adopted description of rotation-induced turbulence in our models, mixing in the tachocline prevents the models from fitting the observed Li evolution in solar-type stars. The efficiency of the turbulent transport in the tachocline as expressed by Eq.~(\ref{eq:DtachBrun}) remains difficult to model correctly as we lack good estimates of the evolution of the ratio $\hat{\Omega}/\Omega$. A fully consistent model should solve the equations for the structure of the turbulent tachocline as derived by \citet[][]{1992A&A...265..106S} with taking into account boundary conditions that describe the variations of the latitudinal differential rotation at the base of the convective envelope as a function of time. This  differential rotation is a function of the global rotation of the star \citep[][]{2017ApJ...836..192B}, and will thus evolve all along its evolution.

 \subsubsection{Parametric turbulence}
 \label{parametricturbulence}
\begin{figure*}[t!]
         \center
         \includegraphics [width=150mm]{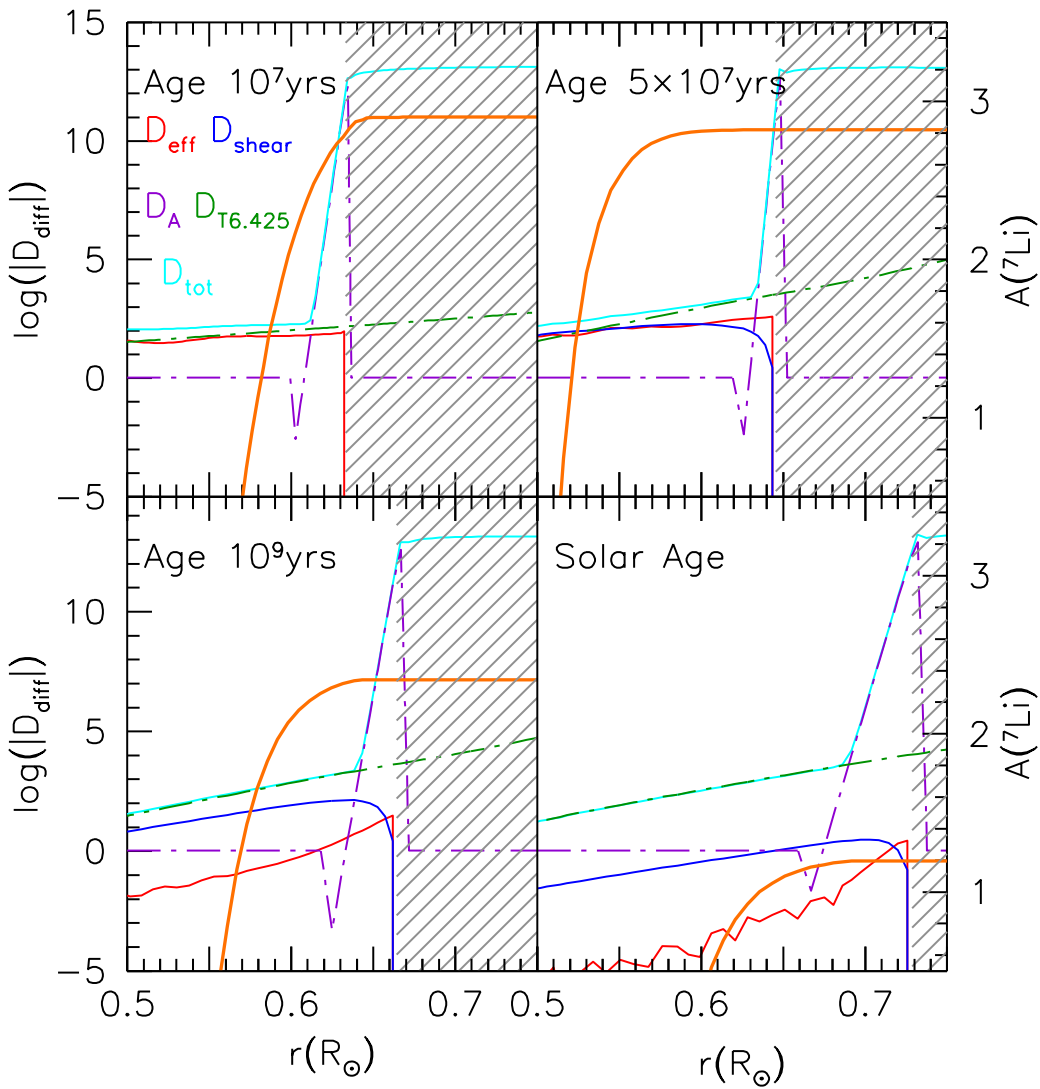}
         \caption{Profiles of the logarithm of the diffusion coefficients (left axis) of the meridional circulation ($D_{\rm{eff}}$), the vertical shear ($D_{\rm{shear}} = D_v$), the overshoot ($D_A$), the parametric turbulence ($D_{T6.425}$), and the total transport coefficient ($D_{tot} = D_{\rm{eff}}+D_{\rm{shear}}+D_A+D_{T6.425}$) as a function of the radius normalised to solar radius at four different ages ($10^7$ years, $5\times10^7$ years, $10^9$ years, and solar age) for model $_{\rm{\nu}}R1_A^{T6.425}$. The abundance profile of $A(^7Li)$ is the orange full line (right axis). Hatched areas correspond to convective regions.}
         \label{fig:prof_coeff}
 \end{figure*}

 Following \citet{2005ApJ...619..538R}, we use the parametric prescriptions for an additional transport of matter as described by Eqs. (\ref{eq:dturb1}) and (\ref{eq:dturb2}). We compute two models, $_{\rm{\nu}}R1_A^{T6.425}$ and $_{\rm{\nu}}R1_A^{PM5000}$, for which we adjust the two free parameters of Eqs. (\ref{eq:dturb1}) and (\ref{eq:dturb2}) as respectively $\log (T_0) = 6.425$ and $a_0 = 5000$ to best fit the evolution of Li abundance observed at the surface of solar-type stars and the Sun. We consider this to be in rather good agreement with the values obtained by \citet[][$\log (T_0) = 6.4$ and $a_0 = 2000$]{2005ApJ...619..538R} for their solar model, considering that it did not include rotation and that it was computed with different basic input physics (in particular nuclear reaction rates, eos, opacities). \\ 
    \begin{figure}[t]
         \center
         \includegraphics [width=90mm]{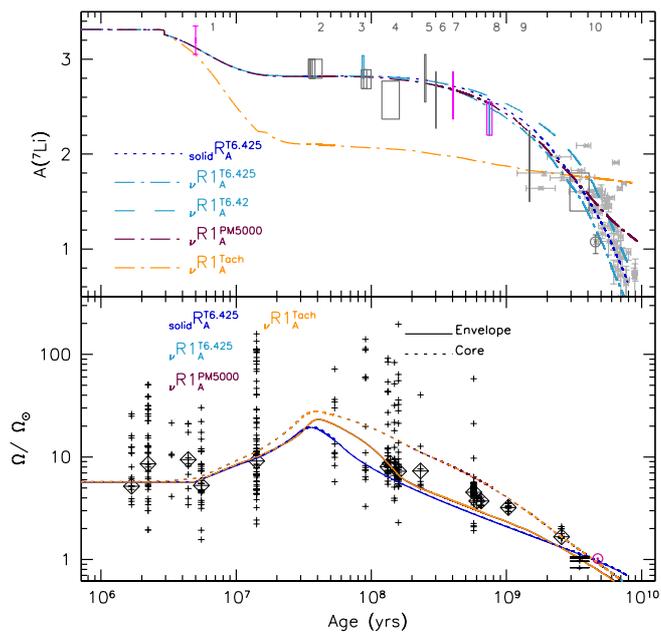}
         \caption{Same as Fig.~\ref{fig:evolLi1}, but for models  $_{\nu}R1^{T6.425}_A$, $_{\nu}R1^{PM5000}_A$, $_{\nu}R1^{Tach}_A$, and $_{\rm{solid}}R1^{T6.425}_A$ that include additional turbulence.}
         \label{fig:evolLi4}
 \end{figure}
  The parametric turbulent mixing becomes efficient beyond 200 - 300 Myr, as can be seen from Fig. \ref{fig:evolLi4}. The form of Eqs. (\ref{eq:dturb1}) and (\ref{eq:dturb2}) leads to an increase in the diffusion coefficient between the base of the convective envelope and the Li burning region as the star evolves on the MS. The evolution of the coefficient $D_{T6.425}$ is illustrated alongside that of the other diffusivities at four different ages in Fig.~\ref{fig:prof_coeff}. We adopt Eq. (\ref{eq:dturb1}) included in model $_{\rm{\nu}}R1_A^{T6.425}$ because it is independent of the depth of the convective envelope and is scaled by the atomic diffusion coefficient for $^4He$ in the Li burning region. While the transport by penetrative convection is the main process responsible for surface Li depletion during the PMS (upper left panel of Fig.~\ref{fig:prof_coeff}), the coefficient $D_{T6.425}$ dominates the transport within the Li burning region on the MS, as shown in the lower panels of Fig.~\ref{fig:prof_coeff} (at 1 Gyr and 4.57 Gyr). The addition of this mixing process does not affect the predicted evolution of the surface rotation rate (lower panel Fig.~\ref{fig:evolLi4}) because it is   independent of rotation. We also tested a solid-body rotating counterpart to model $_{\rm{\nu}}R1_A^{T6.425}$, called model $_{\rm solid}R1_A^{T6.425}$, which is also presented in Fig.~\ref{fig:evolLi4}. It predicts slightly less depletion of Li due to a lower efficiency of the transport in that case. \\ Model $_{\rm{\nu}}R1_A^{T6.425}$ meets the goal of reproducing the surface evolution of the Li abundance and the angular velocity  together with an angular velocity profile compatible with that of the Sun at the solar age.
  
 \subsection{Models with optimal prescriptions} 
 \label{sub:optimalprecript}
 
\begin{figure}[t!]
         \center
         \includegraphics [width=90mm]{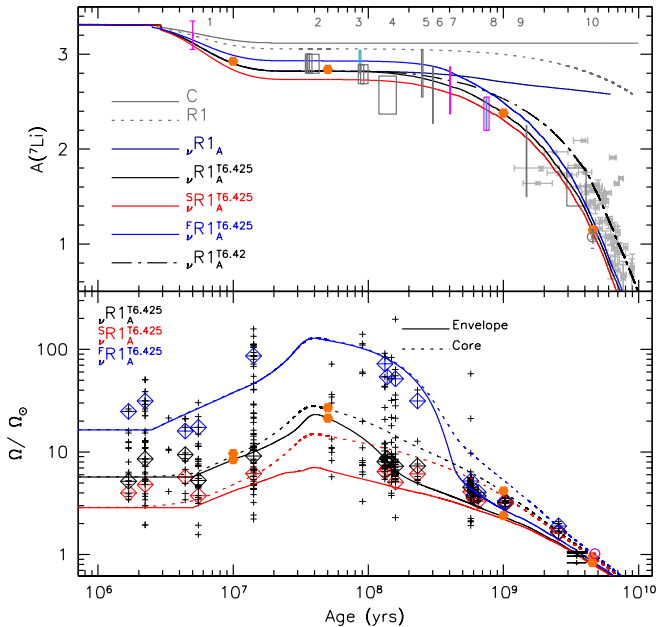}
         \caption{Same as Fig.~\ref{fig:evolLi1}, but  for the different R1 models, the selected $_{\rm{\nu}}R1_A^{T6.425}$ model with different initial rotation velocities, and the classical model (C). The orange dots refer to the ages at which the diffusion coefficient profiles are shown in Fig.\ref{fig:prof_coeff}. The red, black, and blue open squares show the $25{th}$, $50{th}$ and $90{th}$ percentiles of the observed rotational distributions in each cluster.}.
         \label{fig:evolLi5}
 \end{figure}

We explore the impact of the initial angular velocity on our optimal model $_{\rm{\nu}}R1_A^{T6.425}$ and discuss its predictions for other chemical constraints.
We compute $^{\textbf{F}}_{\rm{\nu}}R1_A^{T6.425}$, $_{\rm{\nu}}R1_A^{T6.425}$, and $^{\textbf{S}}_{\rm{\nu}}R1_A^{T6.425}$ for  fast, median, and slow rotators, respectively, and present the associated evolution of Li and surface angular velocity in Fig. \ref{fig:evolLi5}, together with the non-rotating model C (Table \ref{tab:solcal}). 

The relation between surface rotation rate and Li abundance in solar-type stars was first observed by \citet{1993AJ....106.1059S} in the Pleiades and has been confirmed by more recent studies in several clusters \citep{2018A&A...613A..63B,2020arXiv200210556A}. The observations indicate that rapidly rotating stars possess higher Li abundances than slowly rotating stars, which is related to the PMS rotational evolution \citep{2008A&A...489L..53B}. In particular, it is related to the difference in the disc lifetime between the fast and slow rotators \citep[e.g.][]{2012A&A...539A..70E} and/or alternately to the correlation between rotation and penetrative convection efficiency \citep[e.g.][]{2017ApJ...845L...6B}. Contrary to the models of \citet{2016A&A...587A.105A,2019A&A...631A..77A} our models comply with the expected behaviour and the PMS Li depletion is larger for slower rotators, as shown in upper panel of Fig.~\ref{fig:evolLi5}. As we use the same treatment for rotational transport and the same disc lifetimes as in \citet{2019A&A...631A..77A}, the inclusion of penetrative convection according to Eq.~(\ref{eq:dkyle}) is clearly shaping the Li evolution prior to the ZAMS.

On the MS, when the parametric turbulence takes over the Li transport, all models converge to the solar Li value since we have adjusted the parametric turbulence to fit the Sun. The tracks follow the lower envelope of the data from solar twins \citep{2019MNRAS.485.4052C,2020MNRAS.492..245C}. The slightly shallower turbulent mixing used in model $_{\rm{\nu}}R1_A^{T6.42}$ permits us to fit these points. 
The parameters we  adopted for turbulence and magnetic braking lead to a good general agreement between the theoretical and the observed Li behaviour for both field and open cluster solar-type stars.

\begin{figure}[t]
         \center
         \includegraphics [width=90mm]{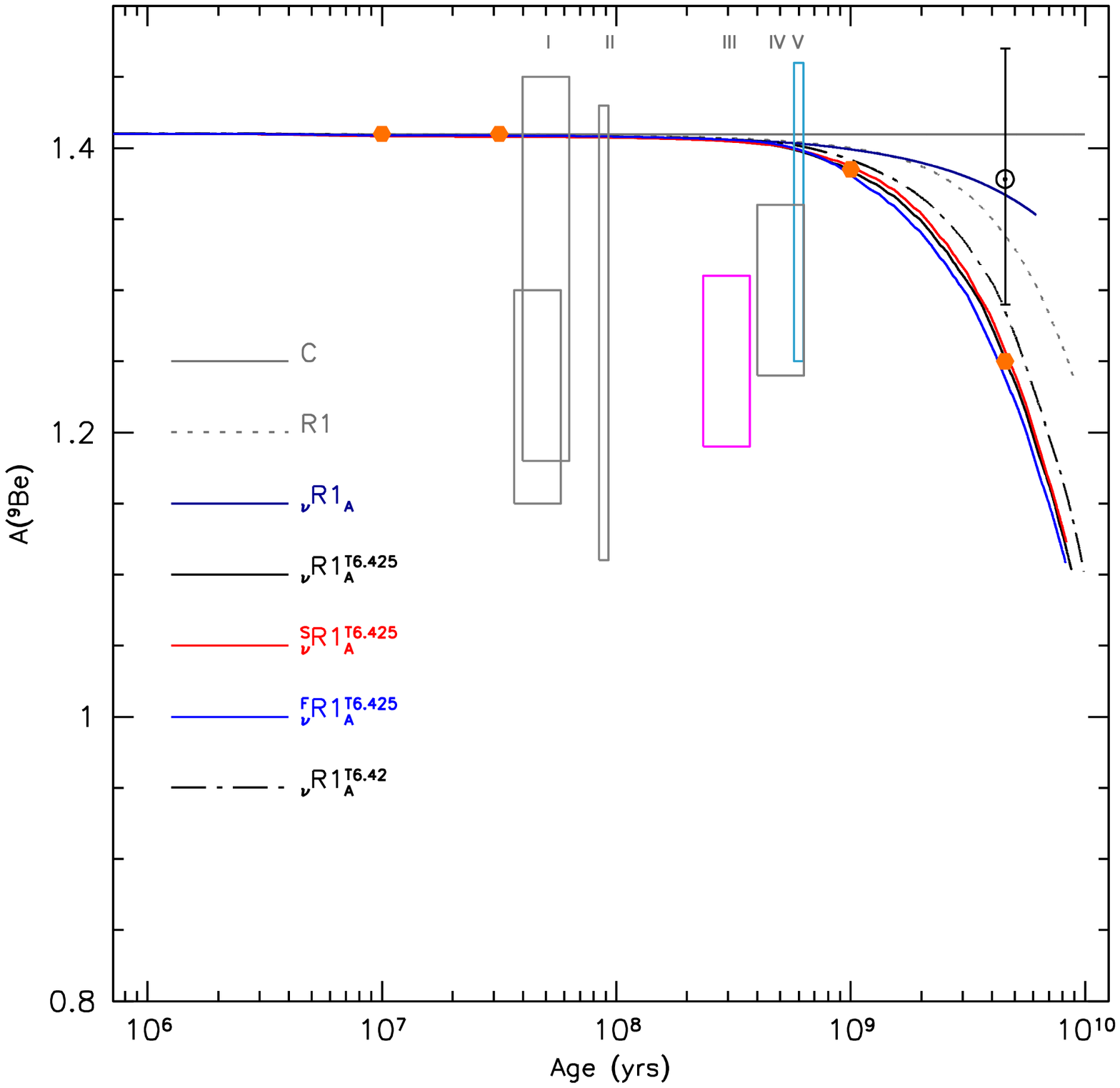}
         \caption{Same as Fig.~\ref{fig:evolLi5},  but for Be. Boxes are for Be observations of solar-mass stars in different open clusters and are colour-coded according to their metallicity, as in Fig.~\ref{fig:evolLi1}. The numbers I to V identify the clusters: I) IC2602 and IC2391 \citep{2011A&A...535A..75S}, II) Pleiades \citep{2003ApJ...583..955B}, III) Ursa Major \citep{2003ApJ...582..410B}, IV) Coma Ber \citep{2003ApJ...582..410B}, and V) Hyades \citep{2004ApJ...605..864B}. The solar beryllium with uncertainties comes from \citet[][photospheric value]{2009ARA&A..47..481A}. The orange dots refer to the four profiles extracted from the star evolution  shown in figure \ref{fig:prof_coeff}.
         }
         \label{fig:evolBe5}
 \end{figure}

Regarding surface and internal rotation, the fast, median, and slow models are also in good agreement with the observed velocity distribution (solid lines, bottom panel Fig.~\ref{fig:evolLi5}) in solar-type stars, similar to the models of \citet{2019A&A...631A..77A}, which did not include additional viscosity for the transport of angular momentum. Moreover, all the models predict a flat rotation profile in the radiative interior at the age of the Sun due to the extra transport of angular momentum by the adopted ad hoc viscosity $\nu_{\rm add}$, regardless of the initial angular velocity. The evolution of the rotation rate  slightly differs, however,  depending on the assumed initial rotational rate.
The model $^F_{\rm{\nu}}R1_A^{T6.425}$ rotates almost as a solid body on the PMS and the early MS. The strong torque exerted by the magnetised winds leads to a sharp deceleration of the surface of this model between 300 Myr and 500 Myr, and differential rotation develops in the interior during that period. When the surface torque becomes inefficient, around 500 Myr, the surface angular velocity evolution settles on a Skumanich-like path \citep[][]{1972ApJ...171..565S} while angular momentum is continuously extracted from the core by the additional transport, modelled here with $\rm{\nu_{add}}$, so that at the age of the Sun the internal rotation profile agrees with the helioseismic constraint. 
On the MS the coupling between the core and the surface is larger for larger initial angular velocity (compare $_{\rm{\nu}}R1_A^{T6.425}$ and $^S_{\rm{\nu}}R1_A^{T6.425}$ in Fig.~\ref{fig:evolLi5}). This behaviour differs from what was obtained from bi-zone models  \citep{2013A&A...556A..36G,2015A&A...584A..30L}. In slow rotators, the large angular velocity gradient and the small surface angular velocity at the ZAMS and during the early MS evolution lead to the stronger Li depletion discussed above.
Beryllium (hereafter Be) is also easily destroyed in stellar interiors, but at higher temperatures ($\approx 3.5$ MK) than Li and can thus also be used to further constrain the mixing.
Spectroscopic determination of Be in open cluster solar-type stars \citep[e.g.][]{2003ApJ...583..955B,2003ApJ...582..410B,2004ApJ...605..864B} indicate that it should be only slightly depleted during the PMS and the MS of solar-like stars. In Fig.~\ref{fig:evolBe5}, we present the evolution of the surface Be abundance as a function of the age for the models shown in Fig.~\ref{fig:evolLi5}.
The evolution of Be at the stellar surface is only affected by the parametric turbulent mixing introduced to reproduce the MS depletion of Li in our models. It does not depend on the initial velocity. The parametric turbulent mixing leads to a 0.3 dex depletion of Be by the age of the Sun, which is slightly too large compared to observations. Model $_{\rm{\nu}}R1_A^{T6.42}$ which better reproduces the Li abundances of solar twins (see previous section), is compatible within the error bars with the Be abundance at the solar surface. However, the limited number of Be abundance determinations currently prevents us from using this nuclide as a good constraint for internal transport processes.

\section{Summary and discussion} 
\label{CONCLUSION}
 We computed models of solar-like stars including atomic diffusion and rotation and we analysed the impact of different internal transport processes on the chemical and rotational evolution of these stars. Our models confirm the need for additional transport processes beyond atomic diffusion and  Type I rotation-induced processes (meridional circulation and turbulent shear) for both angular momentum and chemicals in order to reproduce observations of Li and internal rotation for MS solar-type stars.
In the framework of our study, we chose to parametrise the action of complex processes and adopt the simpler approach of an additional turbulent viscosity $\nu_{\rm{add}}$ to the angular momentum transport equation, either constant or time-dependent. We follow the propositions by \citet{2012A&A...544L...4E} and \citet{2016A&A...589A..23S} to account for the low degree of (radial) differential rotation in evolved stars (subgiants and red giant stars). We show that such a parametrisation of a yet-to-be-identified physical process can indeed lead to a strong coupling between the interior and the envelope in solar-type stars at the age of the Sun without significantly modifying the predicted evolution of their surface rotation. Our models also indicate that all the current observational constraints can be satisfied without reaching a full rotational coupling between the core and the envelope until 2 -3 Gyrs. Moreover, a higher degree of differential rotation is predicted for the slow rotators before this age. Asteroseismic constraints on the internal rotation of young solar-type MS stars would be of great value to test this prediction. 

Concerning the transport of chemicals, we confirm that the Li depletion observed in MS solar-type stars of open clusters cannot be reproduced by rotational mixing alone when the solar internal rotation constraint is also taken into account.
We implemented for the first time the prescription for penetrative convection dependent on rotation, according to \citet{2019ApJ...874...83A}. We show that this process is key to reproducing the surface Li abundance evolution of solar-type stars during the PMS and the very early MS. The dependence of the convective penetration depth on the rotation rate in the convective envelope is the main factor explaining the anti-correlation between the surface rotation rate and Li abundance observed in open clusters. It dominates over the effect of the disc lifetime first pointed out by \citet{2012A&A...539A..70E}.

For the first time since its publication we were able to self-consistently compute the tachocline thickness according to the model by \citet[][their Eq.~(\ref{eq:dtachot})]{1992A&A...265..106S}.
Using the parametrisation for the horizontal turbulent viscosity from \citet{2018A&A...620A..22M}, our model predicts a tachocline thickness at the solar age, which is compatible with the helioseismic estimate by \citet{1999ApJ...516..475E}. On the other hand, the associated turbulent transport as described in \citet{1999ApJ...525.1032B} depends on the degree of latitudinal differential rotation, whose evolution remains poorly constrained. Using estimates based on activity indicators from \citet{1996ApJ...466..384D} leads to very efficient tachocline mixing and an over-depletion of Li surface abundance during the PMS evolution. In addition, the shallow thickness of the tachocline during the MS evolution prevents our models from reproducing the expected Li depletion during this evolutionary phase.
  Furthermore, the models that account self-consistently for the tachocline mixing according to the formalisms of \citet{1992A&A...265..106S} and \citet{1999ApJ...525.1032B} fail to reproduce the observed evolution of Li abundance in solar-type stars. However, this specific formalism calls for observational constraints on the degree of latitudinal differential rotation. The development of the meridional circulation to the fourth order, as proposed by \citet{2004A&A...425..229M}, was designed to self-consistently include the tachocline region and associated transport in stellar evolution models and could help to sort out the actual impact of tachocline mixing along the evolution of solar-type stars. In this context new developments on the horizontal turbulent transport induced by the instability of the horizontal shear of the differential rotation would be of great importance \citep[][]{2020A&A...635A.133P,2020ApJ...901..146G,2020JFM...903A...1C}.\\
  
 Our optimal models ($^X_{\rm{\nu}}R1_A^{T6.425}$) are obtained when including a parametric turbulence according to \citet{2000ApJ...529..338R}, which was first proposed as a way to counteract the impact of atomic diffusion with radiative accelerations in F-, A-, and B-type stars for several evolutionary stages. The depth of the turbulence appears to be slightly greater for the Sun than for solar twins based on the comparison of our models ($_{\rm{\nu}}R1_A^{T6.425}$ and $_{\rm{\nu}}R1_A^{T6.42}$) to the Li and Be data in open clusters, confirming that the Sun might not be the best benchmark for testing transport processes in solar-type stars \citep[e.g. see also][]{2020MNRAS.492..245C}.
\begin{acknowledgements}
 We thank J.W. Ferguson for providing us with low-temperature opacity tables adapted to the solar mixture we adopt in this work, P. Eggenberger for fruitful discussions, and the anonymous referee for constructive comments on the manuscript. 
    This work was supported by the Swiss National Science Foundation (Projects 200021-169125 and 200020-192039 PI C.C.). We thank the Programme National de Physique Stellaire (PNPS) of CNRS/INSU co-funded by CEA and CNES. This research has made use of NASA?s Astrophysics Data System Bibliographic Services. LA acknowledges funding from the European Research Council (ERC) under the European Union's Horizon 2020 research and innovation program (grant agreement No. 682393 AWESoMe-Stars). KCA and SM acknowledge support from the ERC SPIRE 647383 grant and PLATO \& GOLF CNES grants at CEA/DAp-AIM. AP acknowledge support from the PLATO CNES grant at LUPM.
\end{acknowledgements}

%
%

\bibliographystyle{aa}
\bibliography{references}

\begin{thebibliography}{194}
\expandafter\ifx\csname natexlab\endcsname\relax\def\natexlab#1{#1}\fi

\bibitem[{{Aerts} {et~al.}(2019){Aerts}, {Mathis}, \&
  {Rogers}}]{2019ARA&A..57...35A}
{Aerts}, C., {Mathis}, S., \& {Rogers}, T.~M. 2019, \araa, 57, 35

\bibitem[{{Allen}(1976)}]{1976asqu.book.....A}
{Allen}, C.~W. 1976, {Astrophysical Quantities}

\bibitem[{{Amard} {et~al.}(2016){Amard}, {Palacios}, {Charbonnel}, {Gallet}, \&
  {Bouvier}}]{2016A&A...587A.105A}
{Amard}, L., {Palacios}, A., {Charbonnel}, C., {Gallet}, F., \& {Bouvier}, J.
  2016, \aap, 587, A105

\bibitem[{{Amard} {et~al.}(2019){Amard}, {Palacios}, {Charbonnel}, {Gallet},
  {Georgy}, {Lagarde}, \& {Siess}}]{2019A&A...631A..77A}
{Amard}, L., {Palacios}, A., {Charbonnel}, C., {et~al.} 2019, \aap, 631, A77

\bibitem[{{Arancibia-Silva} {et~al.}(2020){Arancibia-Silva}, {Bouvier}, {Bayo},
  {Galli}, {Brandner}, {Bouy}, \& {Barrado}}]{2020arXiv200210556A}
{Arancibia-Silva}, J., {Bouvier}, J., {Bayo}, A., {et~al.} 2020, arXiv
  e-prints, arXiv:2002.10556

\bibitem[{{Asplund} {et~al.}(2009){Asplund}, {Grevesse}, {Sauval}, \&
  {Scott}}]{2009ARA&A..47..481A}
{Asplund}, M., {Grevesse}, N., {Sauval}, A.~J., \& {Scott}, P. 2009, \araa, 47,
  481

\bibitem[{{Augustson} {et~al.}(2012){Augustson}, {Brown}, {Brun}, {Miesch}, \&
  {Toomre}}]{2012ApJ...756..169A}
{Augustson}, K.~C., {Brown}, B.~P., {Brun}, A.~S., {Miesch}, M.~S., \&
  {Toomre}, J. 2012, \apj, 756, 169

\bibitem[{{Augustson} \& {Mathis}(2019)}]{2019ApJ...874...83A}
{Augustson}, K.~C. \& {Mathis}, S. 2019, \apj, 874, 83

\bibitem[{{Augustson} \& {Mathis}(2020)}]{2020svos.conf..311A}
{Augustson}, K.~C. \& {Mathis}, S. 2020, in Stars and their Variability
  Observed from Space, ed. C.~{Neiner}, W.~W. {Weiss}, D.~{Baade}, R.~E.
  {Griffin}, C.~C. {Lovekin}, \& A.~F.~J. {Moffat}, 311--312

\bibitem[{{Baglin} \& {Lebreton}(1990)}]{1990ASSL..159..437B}
{Baglin}, A. \& {Lebreton}, Y. 1990, Astrophysics and Space Science Library,
  Vol. 159, {Surface Abundances of Light Elements as Diagnostic of Transport
  Processes in the Sun and Solar-Type Stars}, ed. G.~{Berthomieu} \&
  M.~{Cribier}, 437

\bibitem[{{Baglin} {et~al.}(1985){Baglin}, {Morel}, \&
  {Schatzman}}]{1985A&A...149..309B}
{Baglin}, A., {Morel}, P.~J., \& {Schatzman}, E. 1985, \aap, 149, 309

\bibitem[{{Bahcall} {et~al.}(1995){Bahcall}, {Pinsonneault}, \&
  {Wasserburg}}]{1995RvMP...67..781B}
{Bahcall}, J.~N., {Pinsonneault}, M.~H., \& {Wasserburg}, G.~J. 1995, Reviews
  of Modern Physics, 67, 781

\bibitem[{{Baraffe} {et~al.}(2017){Baraffe}, {Pratt}, {Goffrey}, {Constantino},
  {Folini}, {Popov}, {Walder}, \& {Viallet}}]{2017ApJ...845L...6B}
{Baraffe}, I., {Pratt}, J., {Goffrey}, T., {et~al.} 2017, \apjl, 845, L6

\bibitem[{{Barker} {et~al.}(2014){Barker}, {Dempsey}, \&
  {Lithwick}}]{2014ApJ...791...13B}
{Barker}, A.~J., {Dempsey}, A.~M., \& {Lithwick}, Y. 2014, \apj, 791, 13

\bibitem[{{Barnes} {et~al.}(2016){Barnes}, {Weingrill}, {Fritzewski},
  {Strassmeier}, \& {Platais}}]{2016ApJ...823...16B}
{Barnes}, S.~A., {Weingrill}, J., {Fritzewski}, D., {Strassmeier}, K.~G., \&
  {Platais}, I. 2016, \apj, 823, 16

\bibitem[{{Basu} \& {Antia}(1995)}]{1995MNRAS.276.1402B}
{Basu}, S. \& {Antia}, H.~M. 1995, \mnras, 276, 1402

\bibitem[{{Beck} {et~al.}(2017){Beck}, {do Nascimento}, {Duarte}, {Salabert},
  {Tkachenko}, {Mathis}, {Mathur}, {Garc{\'\i}a}, {Castro}, {Pall{\'e}},
  {Egeland }, {Montes}, {Creevey}, {Andersen}, {Kamath}, \& {van
  Winckel}}]{2017A&A...602A..63B}
{Beck}, P.~G., {do Nascimento}, J.~D., J., {Duarte}, T., {et~al.} 2017, \aap,
  602, A63

\bibitem[{{Belkacem} {et~al.}(2015){Belkacem}, {Marques}, {Goupil}, {Mosser},
  {Sonoi}, {Ouazzani}, {Dupret}, {Mathis}, \& {Grosjean}}]{2015A&A...579A..31B}
{Belkacem}, K., {Marques}, J.~P., {Goupil}, M.~J., {et~al.} 2015, \aap, 579,
  A31

\bibitem[{{Benomar} {et~al.}(2015){Benomar}, {Takata}, {Shibahashi},
  {Ceillier}, \& {Garc{\'\i}a}}]{2015MNRAS.452.2654B}
{Benomar}, O., {Takata}, M., {Shibahashi}, H., {Ceillier}, T., \&
  {Garc{\'\i}a}, R.~A. 2015, \mnras, 452, 2654

\bibitem[{{Boesgaard}(1976)}]{1976PASP...88..353B}
{Boesgaard}, A.~M. 1976, \pasp, 88, 353

\bibitem[{{Boesgaard}(1991)}]{1991ApJ...370L..95B}
{Boesgaard}, A.~M. 1991, \apjl, 370, L95

\bibitem[{{Boesgaard} {et~al.}(2003{\natexlab{a}}){Boesgaard}, {Armengaud}, \&
  {King}}]{2003ApJ...583..955B}
{Boesgaard}, A.~M., {Armengaud}, E., \& {King}, J.~R. 2003{\natexlab{a}}, \apj,
  583, 955

\bibitem[{{Boesgaard} {et~al.}(2003{\natexlab{b}}){Boesgaard}, {Armengaud}, \&
  {King}}]{2003ApJ...582..410B}
{Boesgaard}, A.~M., {Armengaud}, E., \& {King}, J.~R. 2003{\natexlab{b}}, \apj,
  582, 410

\bibitem[{{Boesgaard} {et~al.}(2004){Boesgaard}, {Armengaud}, \&
  {King}}]{2004ApJ...605..864B}
{Boesgaard}, A.~M., {Armengaud}, E., \& {King}, J.~R. 2004, \apj, 605, 864

\bibitem[{{B{\"o}hm}(1963)}]{1963ApJ...138..297B}
{B{\"o}hm}, K.-H. 1963, \apj, 138, 297

\bibitem[{{B{\"o}hm-Vitense}(1958)}]{1958ZA.....46..108B}
{B{\"o}hm-Vitense}, E. 1958, \zap, 46, 108

\bibitem[{{Bossini} {et~al.}(2019){Bossini}, {Vallenari}, {Bragaglia},
  {Cantat-Gaudin}, {Sordo}, {Balaguer-N{\'u}{\~n}ez}, {Jordi}, {Moitinho},
  {Soubiran}, {Casamiquela}, {Carrera}, \& {Heiter}}]{2019A&A...623A.108B}
{Bossini}, D., {Vallenari}, A., {Bragaglia}, A., {et~al.} 2019, \aap, 623, A108

\bibitem[{{Bouvier}(2008)}]{2008A&A...489L..53B}
{Bouvier}, J. 2008, \aap, 489, L53

\bibitem[{{Bouvier} {et~al.}(2018){Bouvier}, {Barrado}, {Moraux}, {Stauffer},
  {Rebull}, {Hillenbrand}, {Bayo}, {Boisse}, {Bouy}, {DiFolco}, {Lillo-Box}, \&
  {Morales Calder{\'o}n}}]{2018A&A...613A..63B}
{Bouvier}, J., {Barrado}, D., {Moraux}, E., {et~al.} 2018, \aap, 613, A63

\bibitem[{{Brummell}(2007)}]{2007IAUS..239..417B}
{Brummell}, N.~H. 2007, in IAU Symposium, Vol. 239, Convection in Astrophysics,
  ed. F.~{Kupka}, I.~{Roxburgh}, \& K.~L. {Chan}, 417--424

\bibitem[{{Brun} {et~al.}(2017){Brun}, {Strugarek}, {Varela}, {Matt},
  {Augustson}, {Emeriau}, {DoCao}, {Brown}, \& {Toomre}}]{2017ApJ...836..192B}
{Brun}, A.~S., {Strugarek}, A., {Varela}, J., {et~al.} 2017, \apj, 836, 192

\bibitem[{{Brun} {et~al.}(1999){Brun}, {Turck-Chi{\`e}ze}, \&
  {Zahn}}]{1999ApJ...525.1032B}
{Brun}, A.~S., {Turck-Chi{\`e}ze}, S., \& {Zahn}, J.~P. 1999, \apj, 525, 1032

\bibitem[{{Cantiello} {et~al.}(2014){Cantiello}, {Mankovich}, {Bildsten},
  {Christensen-Dalsgaard}, \& {Paxton}}]{2014ApJ...788...93C}
{Cantiello}, M., {Mankovich}, C., {Bildsten}, L., {Christensen-Dalsgaard}, J.,
  \& {Paxton}, B. 2014, \apj, 788, 93

\bibitem[{{Carlos} {et~al.}(2020){Carlos}, {Mel{\'e}ndez}, {do Nascimento}, \&
  {Castro}}]{2020MNRAS.492..245C}
{Carlos}, M., {Mel{\'e}ndez}, J., {do Nascimento}, J.-D., \& {Castro}, M. 2020,
  \mnras, 492, 245

\bibitem[{{Carlos} {et~al.}(2019){Carlos}, {Mel{\'e}ndez}, {Spina}, {dos
  Santos}, {Bedell}, {Ramirez}, {Asplund}, {Bean}, {Yong}, {Yana Galarza}, \&
  {Alves-Brito}}]{2019MNRAS.485.4052C}
{Carlos}, M., {Mel{\'e}ndez}, J., {Spina}, L., {et~al.} 2019, \mnras, 485, 4052

\bibitem[{{Castro} {et~al.}(2009){Castro}, {Vauclair}, {Richard}, \&
  {Santos}}]{2009A&A...494..663C}
{Castro}, M., {Vauclair}, S., {Richard}, O., \& {Santos}, N.~C. 2009, \aap,
  494, 663

\bibitem[{{Ceillier} {et~al.}(2013){Ceillier}, {Eggenberger}, {Garc{\'\i}a}, \&
  {Mathis}}]{2013A&A...555A..54C}
{Ceillier}, T., {Eggenberger}, P., {Garc{\'\i}a}, R.~A., \& {Mathis}, S. 2013,
  \aap, 555, A54

\bibitem[{{Chaboyer} \& {Zahn}(1992)}]{1992A&A...253..173C}
{Chaboyer}, B. \& {Zahn}, J.~P. 1992, \aap, 253, 173

\bibitem[{{Charbonnel} {et~al.}(2013){Charbonnel}, {Decressin}, {Amard},
  {Palacios}, \& {Talon}}]{2013A&A...554A..40C}
{Charbonnel}, C., {Decressin}, T., {Amard}, L., {Palacios}, A., \& {Talon}, S.
  2013, \aap, 554, A40

\bibitem[{{Charbonnel} \& {Talon}(2005)}]{2005Sci...309.2189C}
{Charbonnel}, C. \& {Talon}, S. 2005, Science, 309, 2189

\bibitem[{{Charbonnel} \& {Talon}(2008)}]{2008IAUS..252..163C}
{Charbonnel}, C. \& {Talon}, S. 2008, in IAU Symposium, Vol. 252, The Art of
  Modeling Stars in the 21st Century, ed. L.~{Deng} \& K.~L. {Chan}, 163--174

\bibitem[{{Charbonnel} {et~al.}(1994){Charbonnel}, {Vauclair}, {Maeder},
  {Meynet}, \& {Schaller}}]{1994A&A...283..155C}
{Charbonnel}, C., {Vauclair}, S., {Maeder}, A., {Meynet}, G., \& {Schaller}, G.
  1994, \aap, 283, 155

\bibitem[{{Charbonnel} {et~al.}(1992){Charbonnel}, {Vauclair}, \&
  {Zahn}}]{1992A&A...255..191C}
{Charbonnel}, C., {Vauclair}, S., \& {Zahn}, J.~P. 1992, \aap, 255, 191

\bibitem[{{Chen} \& {Zhao}(2006)}]{2006AJ....131.1816C}
{Chen}, Y.~Q. \& {Zhao}, G. 2006, \aj, 131, 1816

\bibitem[{{Christensen-Dalsgaard} \& {Schou}(1988)}]{1988ESASP.286..149C}
{Christensen-Dalsgaard}, J. \& {Schou}, J. 1988, in ESA Special Publication,
  Vol. 286, Seismology of the Sun and Sun-Like Stars, ed. E.~J. {Rolfe},
  149--153

\bibitem[{{Cope} {et~al.}(2020){Cope}, {Garaud}, \&
  {Caulfield}}]{2020JFM...903A...1C}
{Cope}, L., {Garaud}, P., \& {Caulfield}, C.~P. 2020, Journal of Fluid
  Mechanics, 903, A1

\bibitem[{{Cox} \& {Giuli}(1968)}]{1968pss..book.....C}
{Cox}, J.~P. \& {Giuli}, R.~T. 1968, {Principles of stellar structure}

\bibitem[{{Cranmer} \& {Saar}(2011)}]{2011ApJ...741...54C}
{Cranmer}, S.~R. \& {Saar}, S.~H. 2011, \apj, 741, 54

\bibitem[{{Cummings} {et~al.}(2017){Cummings}, {Deliyannis}, {Maderak}, \&
  {Steinhauer}}]{2017AJ....153..128C}
{Cummings}, J.~D., {Deliyannis}, C.~P., {Maderak}, R.~M., \& {Steinhauer}, A.
  2017, \aj, 153, 128

\bibitem[{{D'Antona} \& {Mazzitelli}(1994)}]{1994ApJS...90..467D}
{D'Antona}, F. \& {Mazzitelli}, I. 1994, \apjs, 90, 467

\bibitem[{{Deal} {et~al.}(2018){Deal}, {Alecian}, {Lebreton}, {Goupil},
  {Marques}, {LeBlanc}, {Morel}, \& {Pichon}}]{2018A&A...618A..10D}
{Deal}, M., {Alecian}, G., {Lebreton}, Y., {et~al.} 2018, \aap, 618, A10

\bibitem[{{Decressin} {et~al.}(2009){Decressin}, {Mathis}, {Palacios}, {Siess},
  {Talon}, {Charbonnel}, \& {Zahn}}]{2009A&A...495..271D}
{Decressin}, T., {Mathis}, S., {Palacios}, A., {et~al.} 2009, \aap, 495, 271

\bibitem[{{Deheuvels} {et~al.}(2015){Deheuvels}, {Ballot}, {Beck}, {Mosser},
  {{\O}stensen}, {Garc{\'\i}a}, \& {Goupil}}]{2015A&A...580A..96D}
{Deheuvels}, S., {Ballot}, J., {Beck}, P.~G., {et~al.} 2015, \aap, 580, A96

\bibitem[{{Deheuvels} {et~al.}(2020){Deheuvels}, {Ballot}, {Eggenberger},
  {Spada}, {Noll}, \& {den Hartogh}}]{2020arXiv200702585D}
{Deheuvels}, S., {Ballot}, J., {Eggenberger}, P., {et~al.} 2020, arXiv
  e-prints, arXiv:2007.02585

\bibitem[{{Deheuvels} {et~al.}(2014){Deheuvels}, {Do{\u{g}}an}, {Goupil},
  {Appourchaux}, {Benomar}, {Bruntt}, {Campante}, {Casagrande}, {Ceillier},
  {Davies}, {De Cat}, {Fu}, {Garc{\'\i}a}, {Lobel}, {Mosser}, {Reese},
  {Regulo}, {Schou}, {Stahn}, {Thygesen}, {Yang}, {Chaplin},
  {Christensen-Dalsgaard}, {Eggenberger}, {Gizon}, {Mathis},
  {Molenda-{\.Z}akowicz}, \& {Pinsonneault}}]{2014A&A...564A..27D}
{Deheuvels}, S., {Do{\u{g}}an}, G., {Goupil}, M.~J., {et~al.} 2014, \aap, 564,
  A27

\bibitem[{{Deheuvels} {et~al.}(2012){Deheuvels}, {Garc{\'\i}a}, {Chaplin},
  {Basu}, {Antia}, {Appourchaux}, {Benomar}, {Davies}, {Elsworth}, {Gizon},
  {Goupil}, {Reese}, {Regulo}, {Schou}, {Stahn}, {Casagrande},
  {Christensen-Dalsgaard}, {Fischer}, {Hekker}, {Kjeldsen}, {Mathur}, {Mosser},
  {Pinsonneault}, {Valenti}, {Christiansen}, {Kinemuchi}, \&
  {Mullally}}]{2012ApJ...756...19D}
{Deheuvels}, S., {Garc{\'\i}a}, R.~A., {Chaplin}, W.~J., {et~al.} 2012, \apj,
  756, 19

\bibitem[{{Delgado Mena} {et~al.}(2014){Delgado Mena}, {Israelian},
  {Gonz{\'a}lez Hern{\'a}ndez}, {Sousa}, {Mortier}, {Santos}, {Adibekyan},
  {Fernand es}, {Rebolo}, {Udry}, \& {Mayor}}]{2014A&A...562A..92D}
{Delgado Mena}, E., {Israelian}, G., {Gonz{\'a}lez Hern{\'a}ndez}, J.~I.,
  {et~al.} 2014, \aap, 562, A92

\bibitem[{{Deliyannis} {et~al.}(2000){Deliyannis}, {Pinsonneault}, \&
  {Charbonnel}}]{2000IAUS..198...61D}
{Deliyannis}, C.~P., {Pinsonneault}, M.~H., \& {Charbonnel}, C. 2000, in IAU
  Symposium, Vol. 198, The Light Elements and their Evolution, ed. L.~{da
  Silva}, R.~{de Medeiros}, \& M.~{Spite}, 61

\bibitem[{{Denissenkov} {et~al.}(2010){Denissenkov}, {Pinsonneault},
  {Terndrup}, \& {Newsham}}]{2010ApJ...716.1269D}
{Denissenkov}, P.~A., {Pinsonneault}, M., {Terndrup}, D.~M., \& {Newsham}, G.
  2010, \apj, 716, 1269

\bibitem[{{Do Nascimento} {et~al.}(2009){Do Nascimento}, {Castro},
  {Mel{\'e}ndez}, {Bazot}, {Th{\'e}ado}, {Porto de Mello}, \& {de
  Medeiros}}]{2009A&A...501..687D}
{Do Nascimento}, J.~D., J., {Castro}, M., {Mel{\'e}ndez}, J., {et~al.} 2009,
  \aap, 501, 687

\bibitem[{{Domingo} {et~al.}(1995){Domingo}, {Fleck}, \&
  {Poland}}]{1995SoPh..162....1D}
{Domingo}, V., {Fleck}, B., \& {Poland}, A.~I. 1995, \solphys, 162, 1

\bibitem[{{Donahue} {et~al.}(1996){Donahue}, {Saar}, \&
  {Baliunas}}]{1996ApJ...466..384D}
{Donahue}, R.~A., {Saar}, S.~H., \& {Baliunas}, S.~L. 1996, \apj, 466, 384

\bibitem[{{dos Santos} {et~al.}(2016){dos Santos}, {Mel{\'e}ndez}, {do
  Nascimento}, {Bedell}, {Ram{\'\i}rez}, {Bean}, {Asplund}, {Spina},
  {Dreizler}, {Alves-Brito}, \& {Casagrande}}]{2016A&A...592A.156D}
{dos Santos}, L.~A., {Mel{\'e}ndez}, J., {do Nascimento}, J.-D., {et~al.} 2016,
  \aap, 592, A156

\bibitem[{{Eddington}(1929)}]{1929MNRAS..90...54E}
{Eddington}, A.~S. 1929, \mnras, 90, 54

\bibitem[{{Eff-Darwich} {et~al.}(2008){Eff-Darwich}, {Korzennik},
  {Jim{\'e}nez-Reyes}, \& {Garc{\'\i}a}}]{2008ApJ...679.1636E}
{Eff-Darwich}, A., {Korzennik}, S.~G., {Jim{\'e}nez-Reyes}, S.~J., \&
  {Garc{\'\i}a}, R.~A. 2008, \apj, 679, 1636

\bibitem[{{Eggenberger} {et~al.}(2019{\natexlab{a}}){Eggenberger}, {Buldgen},
  \& {Salmon}}]{2019A&A...626L...1E}
{Eggenberger}, P., {Buldgen}, G., \& {Salmon}, S.~J.~A.~J. 2019{\natexlab{a}},
  \aap, 626, L1

\bibitem[{{Eggenberger} {et~al.}(2019{\natexlab{b}}){Eggenberger}, {Deheuvels},
  {Miglio}, {Ekstr{\"o}m}, {Georgy}, {Meynet}, {Lagarde}, {Salmon}, {Buldgen},
  {Montalb{\'a}n}, {Spada}, \& {Ballot}}]{2019A&A...621A..66E}
{Eggenberger}, P., {Deheuvels}, S., {Miglio}, A., {et~al.} 2019{\natexlab{b}},
  Astronomy and Astrophysics, 621, A66

\bibitem[{{Eggenberger} {et~al.}(2019{\natexlab{c}}){Eggenberger}, {den
  Hartogh}, {Buldgen}, {Meynet}, {Salmon}, \&
  {Deheuvels}}]{2019A&A...631L...6E}
{Eggenberger}, P., {den Hartogh}, J.~W., {Buldgen}, G., {et~al.}
  2019{\natexlab{c}}, \aap, 631, L6

\bibitem[{{Eggenberger} {et~al.}(2012{\natexlab{a}}){Eggenberger},
  {Haemmerl{\'e}}, {Meynet}, \& {Maeder}}]{2012A&A...539A..70E}
{Eggenberger}, P., {Haemmerl{\'e}}, L., {Meynet}, G., \& {Maeder}, A.
  2012{\natexlab{a}}, \aap, 539, A70

\bibitem[{{Eggenberger} {et~al.}(2017){Eggenberger}, {Lagarde}, {Miglio},
  {Montalb{\'a}n}, {Ekstr{\"o}m}, {Georgy}, {Meynet}, {Salmon}, {Ceillier},
  {Garc{\'\i}a}, {Mathis}, {Deheuvels}, {Maeder}, {den Hartogh}, \&
  {Hirschi}}]{2017A&A...599A..18E}
{Eggenberger}, P., {Lagarde}, N., {Miglio}, A., {et~al.} 2017, \aap, 599, A18

\bibitem[{{Eggenberger} {et~al.}(2005){Eggenberger}, {Maeder}, \&
  {Meynet}}]{2005A&A...440L...9E}
{Eggenberger}, P., {Maeder}, A., \& {Meynet}, G. 2005, \aap, 440, L9

\bibitem[{{Eggenberger} {et~al.}(2008){Eggenberger}, {Meynet}, {Maeder},
  {Hirschi}, {Charbonnel}, {Talon}, \& {Ekstr{\"o}m}}]{2008Ap&SS.316...43E}
{Eggenberger}, P., {Meynet}, G., {Maeder}, A., {et~al.} 2008, \apss, 316, 43

\bibitem[{{Eggenberger} {et~al.}(2010){Eggenberger}, {Meynet}, {Maeder},
  {Miglio}, {Montalban}, {Carrier}, {Mathis}, {Charbonnel}, \&
  {Talon}}]{2010A&A...519A.116E}
{Eggenberger}, P., {Meynet}, G., {Maeder}, A., {et~al.} 2010, \aap, 519, A116

\bibitem[{{Eggenberger} {et~al.}(2012{\natexlab{b}}){Eggenberger},
  {Montalb{\'a}n}, \& {Miglio}}]{2012A&A...544L...4E}
{Eggenberger}, P., {Montalb{\'a}n}, J., \& {Miglio}, A. 2012{\natexlab{b}},
  Astronomy and Astrophysics, 544, L4

\bibitem[{{Eggleton} {et~al.}(1973){Eggleton}, {Faulkner}, \&
  {Flannery}}]{1973A&A....23..325E}
{Eggleton}, P.~P., {Faulkner}, J., \& {Flannery}, B.~P. 1973, \aap, 23, 325

\bibitem[{{Ekstr{\"o}m} {et~al.}(2012){Ekstr{\"o}m}, {Georgy}, {Eggenberger},
  {Meynet}, {Mowlavi}, {Wyttenbach}, {Granada}, {Decressin}, {Hirschi},
  {Frischknecht}, {Charbonnel}, \& {Maeder}}]{2012A&A...537A.146E}
{Ekstr{\"o}m}, S., {Georgy}, C., {Eggenberger}, P., {et~al.} 2012, \aap, 537,
  A146

\bibitem[{{Elliott} \& {Gough}(1999)}]{1999ApJ...516..475E}
{Elliott}, J.~R. \& {Gough}, D.~O. 1999, \apj, 516, 475

\bibitem[{{Elsworth} {et~al.}(1995){Elsworth}, {Howe}, {Isaak}, {McLeod},
  {Miller}, {New}, {Wheeler}, \& {Gough}}]{1995Natur.376..669E}
{Elsworth}, Y., {Howe}, R., {Isaak}, G.~R., {et~al.} 1995, \nat, 376, 669

\bibitem[{{Fuller} {et~al.}(2014){Fuller}, {Lecoanet}, {Cantiello}, \&
  {Brown}}]{2014ApJ...796...17F}
{Fuller}, J., {Lecoanet}, D., {Cantiello}, M., \& {Brown}, B. 2014, \apj, 796,
  17

\bibitem[{{Fuller} {et~al.}(2019){Fuller}, {Piro}, \&
  {Jermyn}}]{2019MNRAS.485.3661F}
{Fuller}, J., {Piro}, A.~L., \& {Jermyn}, A.~S. 2019, \mnras, 485, 3661

\bibitem[{{Gabriel} {et~al.}(1995){Gabriel}, {Grec}, {Charra}, {Robillot},
  {Roca Cort{\'e}s}, {Turck-Chi{\`e}ze}, {Bocchia}, {Boumier}, {Cantin},
  {Cesp{\'e}des}, {Cougrand }, {Cr{\'e}tolle}, {Dam{\'e}}, {Decaudin},
  {Delache}, {Denis}, {Duc}, {Dzitko}, {Fossat}, {Fourmond}, {Garc{\'\i}a},
  {Gough}, {Grivel}, {Herreros}, {Lagard{\`e}re}, {Moalic}, {Pall{\'e}},
  {P{\'e}trou}, {Sanchez}, {Ulrich}, \& {van der Raay}}]{1995SoPh..162...61G}
{Gabriel}, A.~H., {Grec}, G., {Charra}, J., {et~al.} 1995, \solphys, 162, 61

\bibitem[{{Gagnier} \& {Garaud}(2018)}]{2018ApJ...862...36G}
{Gagnier}, D. \& {Garaud}, P. 2018, \apj, 862, 36

\bibitem[{{Gallet} \& {Bouvier}(2013)}]{2013A&A...556A..36G}
{Gallet}, F. \& {Bouvier}, J. 2013, \aap, 556, A36

\bibitem[{{Gallet} \& {Bouvier}(2015)}]{2015A&A...577A..98G}
{Gallet}, F. \& {Bouvier}, J. 2015, \aap, 577, A98

\bibitem[{{Gallet} {et~al.}(2019){Gallet}, {Zanni}, \&
  {Amard}}]{2019A&A...632A...6G}
{Gallet}, F., {Zanni}, C., \& {Amard}, L. 2019, \aap, 632, A6

\bibitem[{{Garaud}(2020)}]{2020ApJ...901..146G}
{Garaud}, P. 2020, \apj, 901, 146

\bibitem[{{Garaud} {et~al.}(2017){Garaud}, {Gagnier}, \&
  {Verhoeven}}]{2017ApJ...837..133G}
{Garaud}, P., {Gagnier}, D., \& {Verhoeven}, J. 2017, \apj, 837, 133

\bibitem[{{Garc{\'\i}a} \& {Ballot}(2019)}]{2019LRSP...16....4G}
{Garc{\'\i}a}, R.~A. \& {Ballot}, J. 2019, Living Reviews in Solar Physics, 16,
  4

\bibitem[{{Garc{\'\i}a} {et~al.}(2014){Garc{\'\i}a}, {Ceillier}, {Salabert},
  {Mathur}, {van Saders}, {Pinsonneault}, {Ballot}, {Beck}, {Bloemen},
  {Campante}, {Davies}, {do Nascimento}, {Mathis}, {Metcalfe}, {Nielsen},
  {Su{\'a}rez}, {Chaplin}, {Jim{\'e}nez}, \& {Karoff}}]{2014A&A...572A..34G}
{Garc{\'\i}a}, R.~A., {Ceillier}, T., {Salabert}, D., {et~al.} 2014, \aap, 572,
  A34

\bibitem[{{Gehan} {et~al.}(2018){Gehan}, {Mosser}, {Michel}, {Samadi}, \&
  {Kallinger}}]{2018A&A...616A..24G}
{Gehan}, C., {Mosser}, B., {Michel}, E., {Samadi}, R., \& {Kallinger}, T. 2018,
  \aap, 616, A24

\bibitem[{{Greenstein} \& {Richardson}(1951)}]{1951ApJ...113..536G}
{Greenstein}, J.~L. \& {Richardson}, R.~S. 1951, \apj, 113, 536

\bibitem[{{Guo} {et~al.}(2017){Guo}, {Lin}, {Bai}, \&
  {Liu}}]{2017Ap&SS.362...15G}
{Guo}, J., {Lin}, L., {Bai}, C., \& {Liu}, J. 2017, \apss, 362, 15

\bibitem[{{Guzik} \& {Mussack}(2010)}]{2010ApJ...713.1108G}
{Guzik}, J.~A. \& {Mussack}, K. 2010, \apj, 713, 1108

\bibitem[{{Harutyunyan} {et~al.}(2018){Harutyunyan}, {Steffen}, {Mott},
  {Caffau}, {Israelian}, {Gonz{\'a}lez Hern{\'a}ndez}, \&
  {Strassmeier}}]{2018A&A...618A..16H}
{Harutyunyan}, G., {Steffen}, M., {Mott}, A., {et~al.} 2018, \aap, 618, A16

\bibitem[{{Hopf}(1930)}]{1930MNRAS..90..287H}
{Hopf}, E. 1930, \mnras, 90, 287

\bibitem[{{Howe} {et~al.}(2020){Howe}, {Chaplin}, {Basu}, {Ball}, {Davies},
  {Elsworth}, {Hale}, {Miglio}, {Nielsen}, \& {Viani}}]{2020MNRAS.493L..49H}
{Howe}, R., {Chaplin}, W.~J., {Basu}, S., {et~al.} 2020, \mnras, 493, L49

\bibitem[{{Iglesias} \& {Rogers}(1996)}]{1996ApJ...464..943I}
{Iglesias}, C.~A. \& {Rogers}, F.~J. 1996, \apj, 464, 943

\bibitem[{{J{\o}rgensen} \& {Weiss}(2018)}]{2018MNRAS.481.4389J}
{J{\o}rgensen}, A. C.~S. \& {Weiss}, A. 2018, \mnras, 481, 4389

\bibitem[{{King} {et~al.}(1997){King}, {Deliyannis}, {Hiltgen}, {Stephens},
  {Cunha}, \& {Boesgaard}}]{1997AJ....113.1871K}
{King}, J.~R., {Deliyannis}, C.~P., {Hiltgen}, D.~D., {et~al.} 1997, \aj, 113,
  1871

\bibitem[{{Korre} {et~al.}(2019){Korre}, {Garaud}, \&
  {Brummell}}]{2019MNRAS.484.1220K}
{Korre}, L., {Garaud}, P., \& {Brummell}, N.~H. 2019, \mnras, 484, 1220

\bibitem[{{Kosovichev}(1988)}]{1988SvAL...14..145K}
{Kosovichev}, A.~G. 1988, Soviet Astronomy Letters, 14, 145

\bibitem[{{Krishna Swamy}(1966)}]{1966ApJ...145..174K}
{Krishna Swamy}, K.~S. 1966, \apj, 145, 174

\bibitem[{{Lagarde} {et~al.}(2012){Lagarde}, {Decressin}, {Charbonnel},
  {Eggenberger}, {Ekstr{\"o}m}, \& {Palacios}}]{2012A&A...543A.108L}
{Lagarde}, N., {Decressin}, T., {Charbonnel}, C., {et~al.} 2012, \aap, 543,
  A108

\bibitem[{{Lagarde} {et~al.}(2014){Lagarde}, {Eggenberger}, {Miglio}, \&
  {Montalb{\`a}n}}]{2014sf2a.conf..483L}
{Lagarde}, N., {Eggenberger}, P., {Miglio}, A., \& {Montalb{\`a}n}, J. 2014, in
  SF2A-2014: Proceedings of the Annual meeting of the French Society of
  Astronomy and Astrophysics, ed. J.~{Ballet}, F.~{Martins}, F.~{Bournaud},
  R.~{Monier}, \& C.~{Reyl{\'e}}, 483--486

\bibitem[{{Lanzafame} \& {Spada}(2015)}]{2015A&A...584A..30L}
{Lanzafame}, A.~C. \& {Spada}, F. 2015, \aap, 584, A30

\bibitem[{{Lebreton} \& {Maeder}(1987)}]{1987A&A...175...99L}
{Lebreton}, Y. \& {Maeder}, A. 1987, \aap, 175, 99

\bibitem[{{Lorenzo-Oliveira} {et~al.}(2020){Lorenzo-Oliveira}, {Mel{\'e}ndez},
  {Ponte}, \& {Galarza}}]{2020MNRAS.tmpL..51L}
{Lorenzo-Oliveira}, D., {Mel{\'e}ndez}, J., {Ponte}, G., \& {Galarza}, J.~Y.
  2020, \mnras [\eprint[arXiv]{2003.13871}]

\bibitem[{{Maeder}(2009)}]{2009pfer.book.....M}
{Maeder}, A. 2009, {Physics, Formation and Evolution of Rotating Stars}

\bibitem[{{Maeder} \& {Zahn}(1998)}]{1998A&A...334.1000M}
{Maeder}, A. \& {Zahn}, J.-P. 1998, \aap, 334, 1000

\bibitem[{{Marques} {et~al.}(2013){Marques}, {Goupil}, {Lebreton}, {Talon},
  {Palacios}, {Belkacem}, {Ouazzani}, {Mosser}, {Moya}, {Morel}, {Pichon},
  {Mathis}, {Zahn}, {Turck-Chi{\`e}ze}, \& {Nghiem}}]{2013A&A...549A..74M}
{Marques}, J.~P., {Goupil}, M.~J., {Lebreton}, Y., {et~al.} 2013, \aap, 549,
  A74

\bibitem[{{Mathis}(2013)}]{2013LNP...865...23M}
{Mathis}, S. 2013, {Transport Processes in Stellar Interiors}, ed. M.~{Goupil},
  K.~{Belkacem}, C.~{Neiner}, F.~{Ligni{\`e}res}, \& J.~J. {Green}, Vol. 865,
  23

\bibitem[{{Mathis} {et~al.}(2004){Mathis}, {Palacios}, \&
  {Zahn}}]{2004A&A...425..243M}
{Mathis}, S., {Palacios}, A., \& {Zahn}, J.~P. 2004, \aap, 425, 243

\bibitem[{{Mathis} {et~al.}(2018){Mathis}, {Prat}, {Amard}, {Charbonnel},
  {Palacios}, {Lagarde}, \& {Eggenberger}}]{2018A&A...620A..22M}
{Mathis}, S., {Prat}, V., {Amard}, L., {et~al.} 2018, Astronomy and
  Astrophysics, 620, A22

\bibitem[{{Mathis} \& {Zahn}(2004)}]{2004A&A...425..229M}
{Mathis}, S. \& {Zahn}, J.~P. 2004, \aap, 425, 229

\bibitem[{{Mathis} \& {Zahn}(2005)}]{2005A&A...440..653M}
{Mathis}, S. \& {Zahn}, J.~P. 2005, \aap, 440, 653

\bibitem[{{Mathur} {et~al.}(2008){Mathur}, {Eff-Darwich}, {Garc{\'\i}a}, \&
  {Turck-Chi{\`e}ze}}]{2008A&A...484..517M}
{Mathur}, S., {Eff-Darwich}, A., {Garc{\'\i}a}, R.~A., \& {Turck-Chi{\`e}ze},
  S. 2008, \aap, 484, 517

\bibitem[{{Matt} {et~al.}(2015){Matt}, {Brun}, {Baraffe}, {Bouvier}, \&
  {Chabrier}}]{2015ApJ...799L..23M}
{Matt}, S.~P., {Brun}, A.~S., {Baraffe}, I., {Bouvier}, J., \& {Chabrier}, G.
  2015, \apj, 799, L23

\bibitem[{{Matt} {et~al.}(2019){Matt}, {Brun}, {Baraffe}, {Bouvier}, \&
  {Chabrier}}]{2019ApJ...870L..27M}
{Matt}, S.~P., {Brun}, A.~S., {Baraffe}, I., {Bouvier}, J., \& {Chabrier}, G.
  2019, \apjl, 870, L27

\bibitem[{{McDonald} \& {Zijlstra}(2015)}]{2015MNRAS.448..502M}
{McDonald}, I. \& {Zijlstra}, A.~A. 2015, \mnras, 448, 502

\bibitem[{{McQuillan} {et~al.}(2014){McQuillan}, {Mazeh}, \&
  {Aigrain}}]{2014ApJS..211...24M}
{McQuillan}, A., {Mazeh}, T., \& {Aigrain}, S. 2014, \apjs, 211, 24

\bibitem[{{Mel{\'e}ndez} \& {Ram{\'\i}rez}(2007)}]{2007ApJ...669L..89M}
{Mel{\'e}ndez}, J. \& {Ram{\'\i}rez}, I. 2007, \apjl, 669, L89

\bibitem[{{Meynet} {et~al.}(2013){Meynet}, {Ekstrom}, {Maeder}, {Eggenberger},
  {Saio}, {Chomienne}, \& {Haemmerl{\'e}}}]{2013LNP...865....3M}
{Meynet}, G., {Ekstrom}, S., {Maeder}, A., {et~al.} 2013, {Models of Rotating
  Massive Stars: Impacts of Various Prescriptions}, ed. M.~{Goupil},
  K.~{Belkacem}, C.~{Neiner}, F.~{Ligni{\`e}res}, \& J.~J. {Green}, Vol. 865, 3

\bibitem[{{Michaud} {et~al.}(2015){Michaud}, {Alecian}, \&
  {Richer}}]{2015ads..book.....M}
{Michaud}, G., {Alecian}, G., \& {Richer}, J. 2015, {Atomic Diffusion in Stars}

\bibitem[{{Montalban}(1994)}]{1994A&A...281..421M}
{Montalban}, J. 1994, \aap, 281, 421

\bibitem[{{Montalb{\'a}n} \& {Rebolo}(2002)}]{2002A&A...386.1039M}
{Montalb{\'a}n}, J. \& {Rebolo}, R. 2002, \aap, 386, 1039

\bibitem[{{Montalban} \& {Schatzman}(1996)}]{1996A&A...305..513M}
{Montalban}, J. \& {Schatzman}, E. 1996, \aap, 305, 513

\bibitem[{{Montalb{\'a}n} \& {Schatzman}(2000)}]{2000A&A...354..943M}
{Montalb{\'a}n}, J. \& {Schatzman}, E. 2000, \aap, 354, 943

\bibitem[{{Morel} {et~al.}(1994){Morel}, {van't Veer}, {Provost}, {Berthomieu},
  {Castelli}, {Cayrel}, {Goupil}, \& {Lebreton}}]{1994A&A...286...91M}
{Morel}, P., {van't Veer}, C., {Provost}, J., {et~al.} 1994, \aap, 286, 91

\bibitem[{{Mosser} {et~al.}(2012){Mosser}, {Goupil}, {Belkacem}, {Marques},
  {Beck}, {Bloemen}, {De Ridder}, {Barban}, {Deheuvels}, {Elsworth}, {Hekker},
  {Kallinger}, {Ouazzani}, {Pinsonneault}, {Samadi}, {Stello}, {Garc{\'\i}a},
  {Klaus}, {Li}, {Mathur}, \& {Morris}}]{2012A&A...548A..10M}
{Mosser}, B., {Goupil}, M.~J., {Belkacem}, K., {et~al.} 2012, \aap, 548, A10

\bibitem[{{Netopil} {et~al.}(2016){Netopil}, {Paunzen}, {Heiter}, \&
  {Soubiran}}]{2016A&A...585A.150N}
{Netopil}, M., {Paunzen}, E., {Heiter}, U., \& {Soubiran}, C. 2016, \aap, 585,
  A150

\bibitem[{{Nielsen} {et~al.}(2014){Nielsen}, {Gizon}, {Schunker}, \&
  {Schou}}]{2014A&A...568L..12N}
{Nielsen}, M.~B., {Gizon}, L., {Schunker}, H., \& {Schou}, J. 2014, \aap, 568,
  L12

\bibitem[{{Palacios} {et~al.}(2006){Palacios}, {Charbonnel}, {Talon}, \&
  {Siess}}]{2006A&A...453..261P}
{Palacios}, A., {Charbonnel}, C., {Talon}, S., \& {Siess}, L. 2006, \aap, 453,
  261

\bibitem[{{Paquette} {et~al.}(1986){Paquette}, {Pelletier}, {Fontaine}, \&
  {Michaud}}]{1986ApJS...61..177P}
{Paquette}, C., {Pelletier}, C., {Fontaine}, G., \& {Michaud}, G. 1986, \apjs,
  61, 177

\bibitem[{{Park} {et~al.}(2020){Park}, {Prat}, \&
  {Mathis}}]{2020A&A...635A.133P}
{Park}, J., {Prat}, V., \& {Mathis}, S. 2020, \aap, 635, A133

\bibitem[{{Piau} \& {Turck-Chi{\`e}ze}(2002)}]{2002ApJ...566..419P}
{Piau}, L. \& {Turck-Chi{\`e}ze}, S. 2002, \apj, 566, 419

\bibitem[{{Pietrinferni} {et~al.}(2013){Pietrinferni}, {Cassisi}, {Salaris}, \&
  {Hidalgo}}]{2013A&A...558A..46P}
{Pietrinferni}, A., {Cassisi}, S., {Salaris}, M., \& {Hidalgo}, S. 2013, \aap,
  558, A46

\bibitem[{{Pin{\c{c}}on} {et~al.}(2017){Pin{\c{c}}on}, {Belkacem}, {Goupil}, \&
  {Marques}}]{2017A&A...605A..31P}
{Pin{\c{c}}on}, C., {Belkacem}, K., {Goupil}, M.~J., \& {Marques}, J.~P. 2017,
  \aap, 605, A31

\bibitem[{{Pinsonneault} {et~al.}(1990){Pinsonneault}, {Kawaler}, \&
  {Demarque}}]{1990ApJS...74..501P}
{Pinsonneault}, M.~H., {Kawaler}, S.~D., \& {Demarque}, P. 1990, \apjs, 74, 501

\bibitem[{{Pinsonneault} {et~al.}(1989){Pinsonneault}, {Kawaler}, {Sofia}, \&
  {Demarque}}]{1989ApJ...338..424P}
{Pinsonneault}, M.~H., {Kawaler}, S.~D., {Sofia}, S., \& {Demarque}, P. 1989,
  \apj, 338, 424

\bibitem[{{Pols} {et~al.}(1995){Pols}, {Tout}, {Eggleton}, \&
  {Han}}]{1995MNRAS.274..964P}
{Pols}, O.~R., {Tout}, C.~A., {Eggleton}, P.~P., \& {Han}, Z. 1995, \mnras,
  274, 964

\bibitem[{{Prat} {et~al.}(2016){Prat}, {Guilet}, {Viallet}, \&
  {M{\"u}ller}}]{2016A&A...592A..59P}
{Prat}, V., {Guilet}, J., {Viallet}, M., \& {M{\"u}ller}, E. 2016, \aap, 592,
  A59

\bibitem[{{Prat} \& {Ligni{\`e}res}(2013)}]{2013A&A...551L...3P}
{Prat}, V. \& {Ligni{\`e}res}, F. 2013, \aap, 551, L3

\bibitem[{{Prat} \& {Ligni{\`e}res}(2014)}]{2014A&A...566A.110P}
{Prat}, V. \& {Ligni{\`e}res}, F. 2014, \aap, 566, A110

\bibitem[{{Pratt} {et~al.}(2017){Pratt}, {Baraffe}, {Goffrey}, {Constantino},
  {Viallet}, {Popov}, {Walder}, \& {Folini}}]{2017A&A...604A.125P}
{Pratt}, J., {Baraffe}, I., {Goffrey}, T., {et~al.} 2017, \aap, 604, A125

\bibitem[{{Proffitt} \& {Michaud}(1991)}]{1991ApJ...380..238P}
{Proffitt}, C.~R. \& {Michaud}, G. 1991, \apj, 380, 238

\bibitem[{{Reimers}(1975)}]{1975MSRSL...8..369R}
{Reimers}, D. 1975, Memoires of the Societe Royale des Sciences de Liege, 8,
  369

\bibitem[{{Richard} {et~al.}(2005){Richard}, {Michaud}, \&
  {Richer}}]{2005ApJ...619..538R}
{Richard}, O., {Michaud}, G., \& {Richer}, J. 2005, The Astrophysical Journal,
  619, 538

\bibitem[{{Richard} {et~al.}(2002){Richard}, {Michaud}, {Richer}, {Turcotte},
  {Turck-Chi{\`e}ze}, \& {VandenBerg}}]{2002ApJ...568..979R}
{Richard}, O., {Michaud}, G., {Richer}, J., {et~al.} 2002, \apj, 568, 979

\bibitem[{{Richard} {et~al.}(1996){Richard}, {Vauclair}, {Charbonnel}, \&
  {Dziembowski}}]{1996A&A...312.1000R}
{Richard}, O., {Vauclair}, S., {Charbonnel}, C., \& {Dziembowski}, W.~A. 1996,
  \aap, 312, 1000

\bibitem[{{Richer} {et~al.}(1998){Richer}, {Michaud}, {Rogers}, {Iglesias},
  {Turcotte}, \& {LeBlanc}}]{1998ApJ...492..833R}
{Richer}, J., {Michaud}, G., {Rogers}, F., {et~al.} 1998, \apj, 492, 833

\bibitem[{{Richer} {et~al.}(2000){Richer}, {Michaud}, \&
  {Turcotte}}]{2000ApJ...529..338R}
{Richer}, J., {Michaud}, G., \& {Turcotte}, S. 2000, The Astrophysical Journal,
  529, 338

\bibitem[{{R{\"u}diger} {et~al.}(2015){R{\"u}diger}, {Gellert}, {Spada}, \&
  {Tereshin}}]{2015A&A...573A..80R}
{R{\"u}diger}, G., {Gellert}, M., {Spada}, F., \& {Tereshin}, I. 2015, \aap,
  573, A80

\bibitem[{{Saar}(2009)}]{2009ASPC..416..375S}
{Saar}, S.~H. 2009, in Astronomical Society of the Pacific Conference Series,
  Vol. 416, Solar-Stellar Dynamos as Revealed by Helio- and Asteroseismology:
  GONG 2008/SOHO 21, ed. M.~{Dikpati}, T.~{Arentoft}, I.~{Gonz{\'a}lez
  Hern{\'a}ndez}, C.~{Lindsey}, \& F.~{Hill}, 375

\bibitem[{{Schatzman}(1993)}]{1993A&A...279..431S}
{Schatzman}, E. 1993, \aap, 279, 431

\bibitem[{{Scherrer} {et~al.}(1995){Scherrer}, {Bogart}, {Bush}, {Hoeksema},
  {Kosovichev}, {Schou}, {Rosenberg}, {Springer}, {Tarbell}, {Title},
  {Wolfson}, {Zayer}, \& {MDI Engineering Team}}]{1995SoPh..162..129S}
{Scherrer}, P.~H., {Bogart}, R.~S., {Bush}, R.~I., {et~al.} 1995, \solphys,
  162, 129

\bibitem[{{Schlattl}(2002)}]{2002A&A...395...85S}
{Schlattl}, H. 2002, \aap, 395, 85

\bibitem[{{Schlattl} \& {Weiss}(1999)}]{1999A&A...347..272S}
{Schlattl}, H. \& {Weiss}, A. 1999, \aap, 347, 272

\bibitem[{{Schwarzschild} {et~al.}(1957){Schwarzschild}, {Howard}, \&
  {H{\"a}rm}}]{1957ApJ...125..233S}
{Schwarzschild}, M., {Howard}, R., \& {H{\"a}rm}, R. 1957, \apj, 125, 233

\bibitem[{{Sestito} \& {Randich}(2005)}]{2005A&A...442..615S}
{Sestito}, P. \& {Randich}, S. 2005, \aap, 442, 615

\bibitem[{{Siess} {et~al.}(2000){Siess}, {Dufour}, \&
  {Forestini}}]{2000A&A...358..593S}
{Siess}, L., {Dufour}, E., \& {Forestini}, M. 2000, \aap, 358, 593

\bibitem[{{Skumanich}(1972)}]{1972ApJ...171..565S}
{Skumanich}, A. 1972, \apj, 171, 565

\bibitem[{{Smiljanic} {et~al.}(2011){Smiljanic}, {Randich}, \&
  {Pasquini}}]{2011A&A...535A..75S}
{Smiljanic}, R., {Randich}, S., \& {Pasquini}, L. 2011, \aap, 535, A75

\bibitem[{{Soderblom} {et~al.}(1993){Soderblom}, {Jones}, {Balachand ran},
  {Stauffer}, {Duncan}, {Fedele}, \& {Hudon}}]{1993AJ....106.1059S}
{Soderblom}, D.~R., {Jones}, B.~F., {Balachand ran}, S., {et~al.} 1993, \aj,
  106, 1059

\bibitem[{{Spada} {et~al.}(2016){Spada}, {Gellert}, {Arlt}, \&
  {Deheuvels}}]{2016A&A...589A..23S}
{Spada}, F., {Gellert}, M., {Arlt}, R., \& {Deheuvels}, S. 2016, \aap, 589, A23

\bibitem[{{Spada} \& {Lanzafame}(2020)}]{2020A&A...636A..76S}
{Spada}, F. \& {Lanzafame}, A.~C. 2020, \aap, 636, A76

\bibitem[{{Spiegel} \& {Zahn}(1992)}]{1992A&A...265..106S}
{Spiegel}, E.~A. \& {Zahn}, J.~P. 1992, \aap, 265, 106

\bibitem[{{Spite} \& {Spite}(1982)}]{1982A&A...115..357S}
{Spite}, F. \& {Spite}, M. 1982, Astronomy and Astrophysics, 115, 357

\bibitem[{{Spruit}(2002)}]{2002A&A...381..923S}
{Spruit}, H.~C. 2002, \aap, 381, 923

\bibitem[{{Stauffer} \& {Hartmann}(1986)}]{1986PASP...98.1233S}
{Stauffer}, J.~B. \& {Hartmann}, L.~W. 1986, \pasp, 98, 1233

\bibitem[{{Takeda} {et~al.}(2010){Takeda}, {Honda}, {Kawanomoto}, {Ando}, \&
  {Sakurai}}]{2010A&A...515A..93T}
{Takeda}, Y., {Honda}, S., {Kawanomoto}, S., {Ando}, H., \& {Sakurai}, T. 2010,
  \aap, 515, A93

\bibitem[{{Takeda} {et~al.}(2007){Takeda}, {Kawanomoto}, {Honda}, {Ando}, \&
  {Sakurai}}]{2007A&A...468..663T}
{Takeda}, Y., {Kawanomoto}, S., {Honda}, S., {Ando}, H., \& {Sakurai}, T. 2007,
  \aap, 468, 663

\bibitem[{{Talon} \& {Charbonnel}(2003)}]{2003A&A...405.1025T}
{Talon}, S. \& {Charbonnel}, C. 2003, \aap, 405, 1025

\bibitem[{{Talon} \& {Charbonnel}(2005)}]{2005A&A...440..981T}
{Talon}, S. \& {Charbonnel}, C. 2005, \aap, 440, 981

\bibitem[{{Talon} \& {Charbonnel}(2010)}]{2010IAUS..268..365T}
{Talon}, S. \& {Charbonnel}, C. 2010, in IAU Symposium, Vol. 268, Light
  Elements in the Universe, ed. C.~{Charbonnel}, M.~{Tosi}, F.~{Primas}, \&
  C.~{Chiappini}, 365--374

\bibitem[{{Talon} {et~al.}(2002){Talon}, {Kumar}, \&
  {Zahn}}]{2002ApJ...574L.175T}
{Talon}, S., {Kumar}, P., \& {Zahn}, J.-P. 2002, \apjl, 574, L175

\bibitem[{{Talon} {et~al.}(2006){Talon}, {Richard}, \&
  {Michaud}}]{2006ApJ...645..634T}
{Talon}, S., {Richard}, O., \& {Michaud}, G. 2006, \apj, 645, 634

\bibitem[{{Talon} \& {Zahn}(1997)}]{1997A&A...317..749T}
{Talon}, S. \& {Zahn}, J.~P. 1997, Astronomy and Astrophysics, 317, 749

\bibitem[{{Th{\'e}venin} {et~al.}(2017){Th{\'e}venin}, {Oreshina}, {Baturin},
  {Gorshkov}, {Morel}, \& {Provost}}]{2017A&A...598A..64T}
{Th{\'e}venin}, F., {Oreshina}, A.~V., {Baturin}, V.~A., {et~al.} 2017, \aap,
  598, A64

\bibitem[{{Thompson} {et~al.}(2003){Thompson}, {Christensen-Dalsgaard},
  {Miesch}, \& {Toomre}}]{2003ARA&A..41..599T}
{Thompson}, M.~J., {Christensen-Dalsgaard}, J., {Miesch}, M.~S., \& {Toomre},
  J. 2003, \araa, 41, 599

\bibitem[{{Thoul} {et~al.}(1994){Thoul}, {Bahcall}, \&
  {Loeb}}]{1994ApJ...421..828T}
{Thoul}, A.~A., {Bahcall}, J.~N., \& {Loeb}, A. 1994, \apj, 421, 828

\bibitem[{{Turcotte} {et~al.}(1998){Turcotte}, {Richer}, {Michaud}, {Iglesias},
  \& {Rogers}}]{1998ApJ...504..539T}
{Turcotte}, S., {Richer}, J., {Michaud}, G., {Iglesias}, C.~A., \& {Rogers},
  F.~J. 1998, \apj, 504, 539

\bibitem[{{Vauclair}(1988)}]{1988ApJ...335..971V}
{Vauclair}, S. 1988, \apj, 335, 971

\bibitem[{{Vauclair}(2013)}]{2013EAS....63..233V}
{Vauclair}, S. 2013, in EAS Publications Series, Vol.~63, EAS Publications
  Series, ed. G.~{Alecian}, Y.~{Lebreton}, O.~{Richard}, \& G.~{Vauclair},
  233--241

\bibitem[{{Vauclair} {et~al.}(1978){Vauclair}, {Vauclair}, {Schatzman}, \&
  {Michaud}}]{1978ApJ...223..567V}
{Vauclair}, S., {Vauclair}, G., {Schatzman}, E., \& {Michaud}, G. 1978, \apj,
  223, 567

\bibitem[{{Waite} {et~al.}(2017){Waite}, {Marsden}, {Carter}, {Petit},
  {Jeffers}, {Morin}, {Vidotto}, {Donati}, \& {BCool
  Collaboration}}]{2017MNRAS.465.2076W}
{Waite}, I.~A., {Marsden}, S.~C., {Carter}, B.~D., {et~al.} 2017, \mnras, 465,
  2076

\bibitem[{{Wallerstein} \& {Conti}(1969)}]{1969ARA&A...7...99W}
{Wallerstein}, G. \& {Conti}, P.~S. 1969, \araa, 7, 99

\bibitem[{{Xing} \& {Xing}(2012)}]{2012A&A...537A..91X}
{Xing}, L.~F. \& {Xing}, Q.~F. 2012, \aap, 537, A91

\bibitem[{{Xu} {et~al.}(2013{\natexlab{a}}){Xu}, {Goriely}, {Jorissen}, {Chen},
  \& {Arnould}}]{2013A&A...549A.106X}
{Xu}, Y., {Goriely}, S., {Jorissen}, A., {Chen}, G.~L., \& {Arnould}, M.
  2013{\natexlab{a}}, \aap, 549, A106

\bibitem[{{Xu} {et~al.}(2013{\natexlab{b}}){Xu}, {Takahashi}, {Goriely},
  {Arnould}, {Ohta}, \& {Utsunomiya}}]{2013NuPhA.918...61X}
{Xu}, Y., {Takahashi}, K., {Goriely}, S., {et~al.} 2013{\natexlab{b}}, \nphysa,
  918, 61

\bibitem[{{Young}(2018)}]{2018ApJ...855...15Y}
{Young}, P.~R. 2018, \apj, 855, 15

\bibitem[{{Zahn}(1991)}]{1991A&A...252..179Z}
{Zahn}, J.~P. 1991, \aap, 252, 179

\bibitem[{{Zahn}(1992)}]{1992A&A...265..115Z}
{Zahn}, J.~P. 1992, \aap, 265, 115

\bibitem[{{Zhang} {et~al.}(2019){Zhang}, {Li}, \&
  {Christensen-Dalsgaard}}]{2019ApJ...881..103Z}
{Zhang}, Q.-S., {Li}, Y., \& {Christensen-Dalsgaard}, J. 2019, \apj, 881, 103

\bibitem[{{Ziegler} \& {R{\"u}diger}(2003)}]{2003A&A...401..433Z}
{Ziegler}, U. \& {R{\"u}diger}, G. 2003, \aap, 401, 433

\end{thebibliography}

\makeatletter
\def\sectcounterend{:}
\def\@seccntformat#1{\appendixname\ \csname the#1\endcsname\sectcounterend%
                      \hskip\betweenumberspace}
\setcounter{section}{0}
\renewcommand\thesection{\@Alph\c@section}
\renewcommand\theequation{\@Alph\c@section.\@arabic\c@equation}
\@addtoreset{equation}{section}
\makeatother

\section{Model convention}
\label{AnnexeA}
In order to simplify the notations, we used the following convention to identify the models:\\
\begin{center}
    \huge{$\rm{^{Rot_{ini}}_{R_{type}}{M_{type}X}^{Turb}_{Oversh}}$},
\end{center}
\normalsize
where we defined
\begin{itemize}
    \item [] Model type ($M_{type}$)
    \begin{itemize}
        \item[.] Classical model (C)
        \item[.] Rotational model (R)
    \end{itemize}
    \item [] Dh/Dv prescription / adjusted K parameter ($X$)
    \begin{itemize}
        \item[.] Dh = \citet{2018A&A...620A..22M} / Dv = \citet{1992A&A...265..115Z} / \\ $K = 7.5\times10^{30}$erg (1)
        \item[.] Dh = \citet{1992A&A...265..115Z} / Dv = \citet{1997A&A...317..749T} / \\ $K = 7.5\times10^{30}$erg (2)
        \item[.] Dh = \citet{1992A&A...265..115Z} / Dv = \citet{1997A&A...317..749T} / \\ $K = 3\times10^{30}$erg (2')
        \item[.] Dh = \citet{1992A&A...265..115Z} / Dv = \citet{1997A&A...317..749T} / \\ $K = 4.5\times10^{30}$erg (2'')
        \item[.] Dh = \citet{2004A&A...425..243M} / Dv = \citet{1992A&A...265..115Z} / \\ $K = 7.5\times10^{30}$erg (3)
    \end{itemize}
    \item [] Initial rotation velocity ($\rm{Rot_{ini}}$)
    \begin{itemize}
        \item[.] Slow (S)
        \item[.] Median (none=default)
        \item[.] Fast (F)
    \end{itemize}
    \item [] Rotation type ($\rm{R_{type}}$) - All rotation models include meridional circulation and shear turbulence
    \begin{itemize}
        \item[.] Impose solid rotation (solid)
        \item[.] Addition of a viscosity $\rm{\nu_{add}}$ ($\nu$)
        \item[.] Addition of the viscosity $\rm{\nu_{add}}$ advised by \citet{2016A&A...589A..23S} for the Sun ($\rm{\nu.spada}$)
        \item[.] Addition of the time-dependent viscosity $\rm{\nu_{add}(t)}$ according to \citet{2016A&A...589A..23S} ($\rm{\nu.spada(t)}$ and $\rm{\nu2.spada(t)}$)
    \end{itemize}
    \item [] Turbulence mixing (Turb)
    \begin{itemize}
        \item[.] Turbulence fixed at temperature T0 (T6.425)
        \item[.] Turbulence fixed at the base of the convection zone (PM5000)
        \item[.] Tachocline turbulence (Tach)
    \end{itemize}
    \item [] Overshoot (Oversh)
    \begin{itemize}
        \item[.] Overshoot from \citet{2017ApJ...845L...6B} (B)
        \item[.] Overshoot from \citet{2019ApJ...874...83A} (A)
        \item[.] Overshoot from \citet{2019MNRAS.484.1220K} (K)
    \end{itemize}
\end{itemize}
\section{Prescriptions for tubulent viscosities}
\label{AnnexeB}
\subsection{Vertical turbulent viscosities}
In the framework of the shellular rotation hypothesis and assuming strong anisotropic turbulence, \textbf{\cite{1992A&A...265..115Z}} proposed  defining  $D_v$ as
\begin{equation}
        D_v = \frac{Ri_{c}}{3} \kappa_T \left(\frac{r~\rm{sin}\theta}{N_T}\frac{d\Omega}{dr}\right)^2,
        \label{eqapp:Zahn92}
 \end{equation}\\
 with $Ri_c$ the critical Richardson number (= 1/4) beyond which the initial instability exists, $\kappa_T$ the thermal diffusivity, $\theta$ a spherical coordinate, and $N_T$ the thermal term of the Brunt-V\"ais\"al\"a frequency.
 \newline
 \\ \textbf{\cite{1997A&A...317..749T}} developed another version of $D_v$ that  considers the effect of thermal and chemical stratifications. The Richardson criterion was then modified; it now involves $N_{\mu}$, the chemical term of the Brunt-V\"ais\"al\"a frequency. The coefficient $D_v$ is defined as
 \begin{equation}
        D_v = \frac{1}{4} Ri_{c} \left(\frac{N_T^2}{K_T+D_h}+\frac{N_{\mu}^2}{D_h}\right)^{-1} \left(r~\rm{sin}\theta \frac{d\Omega}{dr}\right)^2,
 \end{equation}
 where $N^2 = N_T^2 + N_{\mu}^2$ is the Brunt-V\"ais\"al\"a frequency.
 \subsection{Horizontal turbulent viscosities}
 \textbf{\cite{1992A&A...265..115Z}} defined $D_h$ as
 \begin{equation}
        D_h = \frac{1}{c_h} r | 2V_2 - \alpha U_2 |,
 \end{equation}
 with $\alpha= \frac{1}{2}\frac{d ln(r^2 \Omega)}{d ln r}$ the shear rate ($\alpha=1$ means uniform rotation), $V_2$ the latitudinal component, and $U_2$ the radial component of the meridional circulation developed to the second order:
 \begin{equation}
        V_2 = \frac{1}{6 \rho} \frac{d(\rho r^2 U_2)}{dr},
\end{equation}  
\begin{equation}        
        U_2 = 
        \frac{5}{\rho r^4 \Omega} \left(\Gamma(m) - \rho \nu_v r^4 \frac{d \Omega}{dr}\right).
 \end{equation}
 Here $\Gamma(m)$ refers to the gain or loss of angular momentum in the isobar enclosing m(r).
 \newline
 \\ \textbf{\cite{2004A&A...425..243M}} defined $D_h$ as
 \begin{equation}       
        D_h = \left(\frac{\beta}{10}\right)^{1/2} (r^2 \Omega)^{1/2} [r | 2V_2 - \alpha U_2 |]^{1/2},
 \end{equation}
 where $\beta$ is a parameter close to $1.5\times10^{-5}$.
 \newline
\\  \textbf{\cite{2018A&A...620A..22M}} accounted for the fact that horizontal turbulence is generated from both horizontal and vertical shears. The coefficient $D_h$ then writes as $D_h=D_{h,h} + D_{h,v}$. The first index means the direction of the transport and the second index refers to the shear that generates the transport. This additional transport ($D_{h,v}$) is active only if the vertical shear does not fulfil the Reynolds criterion ($R_e > R_{e;c}$ with $R_{e;c} = 7 \nu_m$ with $\nu_m$ the molecular viscosity):
 \begin{equation}       
        D_{h,h} = \left(\frac{\beta}{10}\right)^{1/2} (r^2 \Omega)^{1/2} [r | 2V_2 - \alpha U_2 |]^{1/2}, 
\end{equation}
\begin{equation}        
        D_{h,v} = \left\{  
                \begin{array}{ll}
                \frac{\tau^2 N^4}{2 \Omega^2}  D_{v,v} & \mbox{if } R_e > R_{e;c} \\
                0 & \mbox{if } R_e < R_{e;c}
                \end{array}
                \right \}
.\end{equation}
 Here $D_{v,v} \equiv D_v$ from Eq.~(\ref{eqapp:Zahn92}), and $\tau$ is a characteristic timescale for the turbulence, taken to be equal to the time characterising the radial shear, $\tau = 1/S$ , where $S = r \sin \theta \partial_r \Omega$, as in \cite{2019A&A...631A..77A}.

\end{document}